\documentclass[aps,twocolumn,superscriptaddress]{revtex4}
\usepackage{graphicx,color,appendix,amssymb,amsmath,hyperref}
\usepackage[british]{babel}

\begin{document}

\title{Critical adsorption of polyelectrolytes onto planar and convex highly charged surfaces: the nonlinear Poisson-Boltzmann approach}

\author{Sidney J. de Carvalho}
\email{sidneyjc@ibilce.unesp.br}
\affiliation{Institute of Biosciences, Letters and Exact Sciences, Sao Paulo State University, 15054-000 Sao Jose do Rio Preto, Brazil}
\affiliation{Institute for Physics \& Astronomy, University of Potsdam, 14476 Potsdam-Golm, Germany}

\author{Ralf Metzler}
\email{rmetzler@uni-potsdam.de}
\affiliation{Institute for Physics \& Astronomy, University of Potsdam, 14476 Potsdam-Golm, Germany}

\author{Andrey G. Cherstvy}
\email{a.cherstvy@gmail.com}
\affiliation{Institute for Physics \& Astronomy, University of Potsdam, 14476 Potsdam-Golm, Germany}

\date{\today}

\begin{abstract}

We study the adsorption-desorption transition of polyelectrolyte chains onto planar, cylindrical and spherical surfaces with arbitrarily high surface charge densities by massive Monte Carlo computer simulations. We examine in detail how the well known scaling relations for the threshold transition---demarcating the adsorbed and desorbed domains of a polyelectrolyte near weakly charged surfaces---are altered for highly charged interfaces. In virtue of high surface potentials and large surface charge densities, the Debye--H\"uckel approximation is often not feasible and the nonlinear Poisson-Boltzmann approach should be implemented. At low salt conditions, for instance, the electrostatic potential from the nonlinear Poisson--Boltzmann equation is \textit{smaller }than the Debye--H\"uckel result, such that the required critical surface charge density for polyelectrolyte adsorption $\sigma_c$  \textit{increases}. The nonlinear relation between the surface charge density and electrostatic potential leads to a sharply increasing critical surface charge density with growing ionic strength, imposing an additional limit to the critical salt concentration above which no polyelectrolyte adsorption occurs at all. We contrast our simulations findings with the known scaling results for weak critical polyelectrolyte adsorption onto oppositely charged surfaces for the three standard geometries. Finally, we discuss some applications of our results for some physical-chemical and biophysical systems.

\end{abstract}

\maketitle

\section{Introduction}
\label{sec-int}

The adsorption of charged polymers or polyelectrolytes (PEs) onto 
oppositely charged planar and curved interfaces \cite{netz03,dobr05, kawa82,dobr08, wink14,khok15} attracted the attention of a large number of theoretical \cite{wieg77, wieg86,muth87, joan91,muth94,muth95,adam96,barr96,linse96,adam03, sens99, ande98,netz99, netz99b,netz00,shaf03, netz14,kier13,fors12, cher06a,cher06b,cher07, cher11, cher12a, cher12b,cher14,cher14c,cher15}, experimental \cite{dech97, schl99,scho03, scho07,volo04,dub01a, dub01b,dub05, dub06,dub07, dub10,dub11,dub13, bord12,diet10,diet12, bord15,yu15, bork14}, and computer simulations \cite{belt91, carv10, carv14,stol02, stol03,stol05, stol06,stol09,stol11, dobr07,hoda08,hoda07, vatt08,muth11, dubi10,chin11,vrie04,kong98, muth02,krama96,mess06, mess09,reddy06, caet13,moli14,bach15} groups over the last decades. Important applications of PE--surface adsorption include paper making \cite{lind91}, surface coating \cite{kawa82,bork14}, metal corrosion inhibition via multilayered adsorption \cite{moeh06, andr10},  flocculation as well as stabilisation of colloidal suspensions \cite{napp83,bork14,pinc91}, formation of polymer--nanoparticle composites \cite{pavi08,golu01,malw03}, PE--protein \cite{dubi10,dub13,ball13} and PE--micelle \cite{dubi94} complexation, to mention but a few. Complexation of PEs with nanoparticles is also employed for water treatment \cite{greg01}, utilising polymer flocculation \cite{greg11} with water-dissolved particles. Multilayered PE formation on surfaces \cite{dech97,borg14} and hollow microcapsules \cite{volo04,gees09} has several technological and biomedical (drug delivery) applications \cite{cohe10,lasc00, boud09,skir11}; see Ref. \cite{cull04} for functioning mechanisms of polymer-drug conjugates.

\begin{figure}\includegraphics[width=7cm]{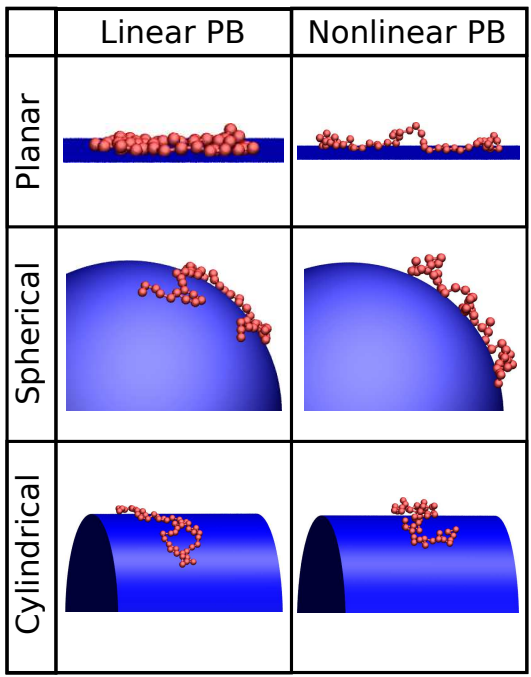}
\caption{Snapshots of typical PE configurations obtained from Monte Carlo simulations for a relatively high surface charge density of $\sigma$=0.5C/m$^2$ and salt concentration of $n_0$=0.7M corresponding to $\kappa\approx0.27$\AA$^{-1}$. ES surface potentials were obtained from solutions of the linear and nonlinear PB equation, see App. \ref{app-potentials}. The cylinder and sphere radius is $a=100$\AA, the chain monomer radius is $R=2$\AA, and the intercharge separation along the PE is $b=7$\AA. At these conditions, the system is close to the adsorption--desorption threshold for the nonlinearly treated ES potential, while for the linear ES potential the PE is in the adsorbed state. Video files illustrating the PE adsorption dynamics for the above mentioned model parameters are presented in the Supplementary Material.}\label{fig-intro}\end{figure}

Both the limits of strong and weak PE--surface adsorption were investigated \cite{wink14}. For the latter, the electrostatic (ES) polymer-surface attraction is comparatively \textit{weak }and the transition between adsorbed and desorbed chain conformations is governed by the interplay of the PE--surface attraction and the entropic penalty of chain confinement in the vicinity of the interface \cite{dobr05,wink14}, see also Fig. \ref{fig-intro}. In this weak coupling limit the transition is quantified in terms of the critical surface charge density $\sigma_c$ via its dependence on the reciprocal Debye screening length, $\kappa$. With increasing solution salinity, the screening of attractive PE--surface interactions becomes stronger, and the surface charge density necessary to get the polymer adsorbed onto the interface from the solution increases. The adsorption--desorption transition of polymers near attractive interfaces is also controlled by the PE charge density, chain bending softness, and temperature \cite{muth87,wink14,wieg77}.

The well known result for a uniformly \textit{weakly charged} planar interface \cite{wieg77}, \begin{equation} \sigma^{l}_{c,pl}(\kappa) \sim \kappa^3, \label{eq-sigma-3} \end{equation} is based on the eigenfunction expansion of the Edwards equation with the Debye--H\"uckel ES potential, see also Refs. \cite{muth87,kong98,cher11,shaf03}. This expansion defines the PE bound states in the attractive potential field of the surface \cite{wieg77, wieg86,muth87, muth94,cher11,cher06b}. In these approaches, the latter is computed from the linear Poisson--Boltzmann (PB) theory and attracts the oppositely charged nearly Gaussian PE chain towards the interface. The scaling relation (\ref{eq-sigma-3}) implies that the statistical properties of PE chains are approximated by those of infinitely long Gaussian neutral polymers \cite{wieg77,cher11,barr96}. 
The analytical modifications of Eq. (\ref{eq-sigma-3}) are known for PE adsorption from dilute solutions onto convex cylindrical and spherical interfaces \cite{cher11,wink14} as well as for PEs under confinement imposed by planar and concave surfaces \cite{cher12a,cher15}. The effects of surface curvature \cite{muth94,netz99b,netz14,cher06b,cher11,cher14}, image forces \cite{mess06, stol09,cher12b}, and chargeable surface groups \cite{stol06} onto PE adsorption were studied within the linear ES\ model. Moreover, some computer simulations results are available for various structured and patterned surfaces \cite{muth95, hoda08, muth02}, spherical Janus particles \cite{cher14}, and PE chains with pH-sensitive charge density \cite{stol05,stol11,carv14}. The implications of the applied shear and hydrodynamic interactions onto PE--surface adsorption were examined as well \cite{hoda07}, see also Ref. \cite{gomp15}. 

In contrast to neutral polymers confined near interfaces \cite{eise93,eise82,eise96}, the adsorption of PE chains onto oppositely charged surface is controlled by
an additional length scale, the Debye  screening length, $\lambda_D=1/\kappa$. Here $\kappa = \sqrt{8 \pi l_B n_0} $ is the reciprocal Debye screening length in symmetric 1:1 electrolyte  with salt concentration $n_0$ and $l_B = e_ 0^2/(\epsilon k_B T) \approx $ 7.1\AA~is the Bjerrum length in the aqueous solution with dielectric constant $\epsilon$=78.7 and at temperature $T = 298.15$K (as used hereafter). At this distance the ES interactions of two unit electric charges are equal to the thermal energy, $k_B T$. This additional length scale not only  dramatically changes the adsorption properties of individual polymer chains, but also affects the ES interactions between the adsorbed PE segments along the charged surface \cite{dobr05,fors12}.

An open question is how different the properties of PE--surface adsorption are for \textit{highly charged} interfaces, when the linear PB approximation is no longer valid \cite{ande95,kapp06,ohsh82}? How strongly the relation (\ref{eq-sigma-3}) will get modified? Both in theoretical and experimental literature the canonical cubic scaling of Eq. (\ref{eq-sigma-3}) is often used as the asymptote in a broad range of parameters, also for situations when the ES potential is not necessarily $\leq$25mV. 

In the current study, we unveil the implications of the nonlinear ES potential distribution emerging near highly charged surfaces \cite{ohsh82, ohsh98} onto the conditions of critical PE adsorption in three basic geometries. The paper is organised as follows. In Sec. \ref{sec-mod} and App. \ref{app-potentials} we describe some details of the derivation of the ES potential and our simulations procedure. The main results are presented in Sec. \ref{sec-res}.  We first consider PE adsorption onto the planar interface with the linear and nonlinear ES potentials, Sec. \ref{sec-planar}. The PE critical adsorption conditions in curved cylindrical and spherical geometries are investigated in Sec. \ref{sec-curved}. The Discussion and Conclusions in Sec. \ref{sec-dis} summarise our findings and provide some applications.

\section{Model and Approximations}
\label{sec-mod}

\subsection{Potential Distribution}

For a planar surface with a charge density $\sigma$ the distribution of the ES potential in symmetric 1:1 electrolyte can be obtained \cite{ande95, kapp06, anti98} from the solution of the nonlinear PB equation
\begin{equation} \nabla^2_\textbf{r} \Psi(\textbf{r})=\kappa^2 \sinh[\Psi(\textbf{r})]. \label{eq-nonlin-pb} \end{equation} Here, the standard dimensionless ES potential is \begin{equation}\Psi=e_0\phi/(k_B T).\label{eq-dimens}\end{equation} The solution of Eq. (\ref{eq-nonlin-pb}) given in App. \ref{app-potentials} is described by Eqs. (\ref{eq-plane-nlpb}), (\ref{eq-nlpb-sphere-final-uni-solution}) and (\ref{eq-solution-uni-cyl}) for planar, spherical and cylindrical surfaces, respectively. The monomer--surface ES interaction energy near the interface is just $e_{0} \phi(\textbf{r}).$ Interactions other than direct  charge--charge forces---e.g. van der Waals forces or electrodynamic fluctuations-induced $r^{-6}$ forces \cite{fren10}---are neglected below. We assume that PE chains in proximity to the surface do not alter the potential distribution emerging from the interface. Near highly charged surfaces, we also neglect possible formation of an immobile Stern layer, structuring of ions, surface charge renormalisation, a lower dielectric permittivity $\epsilon$ due to hydration layers, as well as an altered dielectric response on the nanoscale \cite{korn07, fren10}. So, we employ the so called primitive model for a structureless electrolyte \cite{podg09, podg13, levy12, roij12}, with no explicit solvent or charges being considered in simulations.

The ES interactions of chain monomers are computed via summing up the pairwise screened Coulomb contributions  \begin{equation} E_\text{rep}(r)=e_0^2e^{-\kappa r}/(\epsilon r). \label{eq-screened-repulsion} \end{equation} This implies low-to-moderate Manning-Oosawa PE charge parameter \cite{mann78, oosa70, gros00}, $\xi=l_B/b$, where $b$ is the inter-bead distance, with each bead carrying one elementary charge, $e_0$. For rather long chains, not too highly charged PEs, and strong screening conditions the repulsion of charges along the chain $E_\text{rep}(r)$ can be approximated by this form.

For comparison, we also use in our simulations the ES potential emerging from the surface computed within the linearised PB theory, given respectively by Eqs. (\ref{lpb-plan}), (\ref{eq-lpb-solution-potential}) and (\ref{eq-sol-lpb-cyl}) of App. \ref{app-potentials} for the three standard adsorption geometries. The ES potentials from the linear PB theory are explicit and easy to implement in simulations  \cite{carv10,cher14,cher15}. The potential from the nonlinear PB approach often requires a numerical solution for an \textit{implicit} $\Psi_s(\sigma)$ dependence, somewhat complicating the simulations. Specifically, to restore the distance variation of the ES potential for the planar, spherical, and cylindrical geometries, we first systematically evaluate  $\Psi_s$ for varying $\sigma$ using Eqs. (\ref{eq-gram-plane}), (\ref{eq-grahame-sphere-3}), and (\ref{eq-grahame-cyl}), respectively.

Note that both the linear and nonlinear PB approaches are mean field theories which neglect ion-ion correlation effects. The latter become particularly important e.g. in the presence of multivalent cations and near highly charged surfaces \cite{netz02,gros00,netz05}. Some effects triggered by the binding of multivalent cations are the charge inversion \cite{gros00} and attraction between likely charged objects such as DNA \cite{pars00, korn07, cher11}, often emerging due to correlated Wigner crystal charge density waves on interacting surfaces \cite{gros00,korn07, levi02,wong03,mess09}. Also, at highly charged surfaces and for the finite-size electrolyte ions, the effects of crowding, space restriction, and steric interactions can be important. This yields an additional source of nonlinearity  in such lattice-gas based models of electrolytes, producing Fermi-like \cite{eige59,ande97,ande06, korn07b, podg09} rather than Boltzmann-like charge density distributions in the Poisson equation. All these additional effects are beyond the scope of the current study. Finally, note that the nonlinear PB theory generally violates the superposition principle and features some inconsistencies for calculations of e.g. the ES free energy of charges in electrolytes \cite{onsa33, korn07}. The linearised PB approach satisfies the superposition, but becomes progressively inaccurate for high charge densities \cite{dese00}. Below, we focus on novel physical effects stemming from a different decay law and altered ES potential magnitude near highly charged surfaces, as prescribed by the nonlinear PB approach, onto critical PE-surface adsorption.

The results for the potential variation with the distance from the surface and the nonlinear coupling of the surface charge density $\sigma$ and surface potential $\Psi_s$ are presented in Fig. \ref{fig-pot} for spherical and cylindrical surfaces. Starting from the same value on the surface $\Psi = \Psi_s$, the nonlinear PB potential varies \textit{much faster }with distance from the surface, as compared to the linear PB result. The solution salinity is fixed here to nearly physiological conditions,   $\kappa \sim 1/$(10\AA). We find also that for smaller surface curvature values $a$  the potential approaches the far field asymptote at shorter distances from the surface. 
On the contrary, for larger radii $a$ the short distance asymptote reproduces quite satisfactorily also the long distance behaviour of the ES potential. The approximations for the nonlinear PB potential near the interface and far away from it are shown in Fig. \ref{fig-pot} as the dashed asymptotes to the exact nonlinear PB results. Note also that the initial faster-than-exponential decrease of the  \textit{nonlinear ES potential} with separation is expected to change also the $\sigma_c (\kappa)$ dependence \cite{cher11,wink14} for strongly charged interfaces, as we show below. 

In contrast to the standard linear relation between the surface potential and the surface charge density, the \textit{generalised Grahame relations }for spherical and cylindrical interfaces \cite{ohsh82} lead to much weaker $\Psi(\sigma)$ dependencies for highly charged surfaces, as compared to the linear case  (\ref{eq-linear-psi-sigma-relation}), see the bottom panel of Fig. \ref{fig-pot}. Thus, a higher ES potential will emerge near the interface within the linear PB model at a given $\sigma$, as compared to the nonlinear PB situation. Clearly, at smaller interface charge densities the relation between the surface potential and the charge density obtained from the linear and nonlinear approaches are close to one another. This regime is presented in the bottom panels of Fig. \ref{fig-pot} as the limit of \textit{large area} per elementary charge on the interface.

\subsection{Simulation Method}

In the current study, we use the same physical model, simulations procedure and PE--surface adsorption criteria as in our recent studies \cite{cher14,cher15,caet13}. In short, the PE chain is described within the bead--spring model, with each monomer being a sphere of radius $R$ carrying the central unit charge $e_{0}$. The spring force constant is chosen to yield the mean bead--bead distance of $b\approx$7\AA$ \approx l_B$. Thus, the counterion condensation would not occur on this weakly charged PE, as the Manning theory predicts \cite{mann78}. ES interactions between the monomers are screened Coulomb potentials (\ref{eq-screened-repulsion}), determining also the ES contribution to the chain persistence  \cite{barr96, wink14, cher14}. In addition, the hard core repulsion acts between polymer beads, accounting for excluded volume effects, particularly for compact PE conformations on interfaces. We carried out extensive Metropolis Monte Carlo computer simulations in the canonical NVT-ensemble, see Refs. \cite{carv10,carv14,cher14,cher15}. Three movements of system components were implemented: (1) random displacement of chain monomers, (2) random displacement of the whole chain, and (3) pivot rotation of a chain part around a randomly chosen monomer \cite{soka88}. The equilibration was reached with $\sim10^7$ configurations and we used $\sim10^8$ configurations per particle to calculate the average quantities.

As discussed in Refs. \cite{carv14,cher14,wink14,cher11}, for a given range of surface charge densities the PE chains undergo discontinuous \textit{transitions }between the adsorbed and desorbed states. The fraction of configurations in the desorbed state increases with decreasing surface charge density due to weaker ES attraction of the chain towards the surface. For a given solution salinity, the critical surface charge density $\sigma_c$ was therefore defined as that one in which half of chain configurations recorded over simulation time are in the adsorbed state. This defines the position of the adsorption--desorption boundary in the plane of the model parameters; similar definition was used in our previous studies  \cite{cher14,cher15}.

Therefore, we perform the computer simulations for \textit{systematically varying }surface charge densities, in order to find the range of $\sigma$ in which the coexistence of adsorbed and desorbed states of the chain takes place. Each simulation is executed with $10^7$ Monte Carlo steps. We identify the range of the surface charge densities in which the fraction of PE configurations in the adsorbed state is between 0.4 and 0.6. To determine the precise value of $\sigma_c$, we are running 10 sets of simulations within this preselected range of $\sigma$ using different random number seeds for each set. For each set, the fraction of configurations in the adsorbed state versus the surface charge density dependence is fitted by a sigmoid function that yields the actual value of $\sigma_c$. In this stage, each simulation is executed with $10^8$ steps. The statistical deviations for $\sigma_c$ values as calculated over 10 independent simulations are typically $\lesssim$1\%. These irregularities are reflected e.g. in the magnitudes of the error bars for $\sigma_c$ which are often smaller that the symbol size in Fig. \ref{fig-sig1}. 

In Fig. \ref{fig-intro} the three basic geometries and typical chain configurations are shown as provided by our computer simulations. Here, typical model parameters are used: the surface charge density is $\sigma =$ 0.5C/m$^2$, the solution salinity is 0.7M, and the PE degree of polymerisation is $N=$ 50, both for the linear and nonlinear ES potentials. Figure \ref{fig-intro} demonstrates that in the case of a nonlinear potential the chain gets adsorbed to the surface to \textit{weaker extent}. Generally, \textit{lower }values obtained for the nonlinear ES potential give rise to \textit{higher }surface charge densities required to trigger critical PE adsorption, as compared to a surface with the linearly treated ES\ potential. The ramifications of this fact is the main subject of the current study.

\begin{figure*}
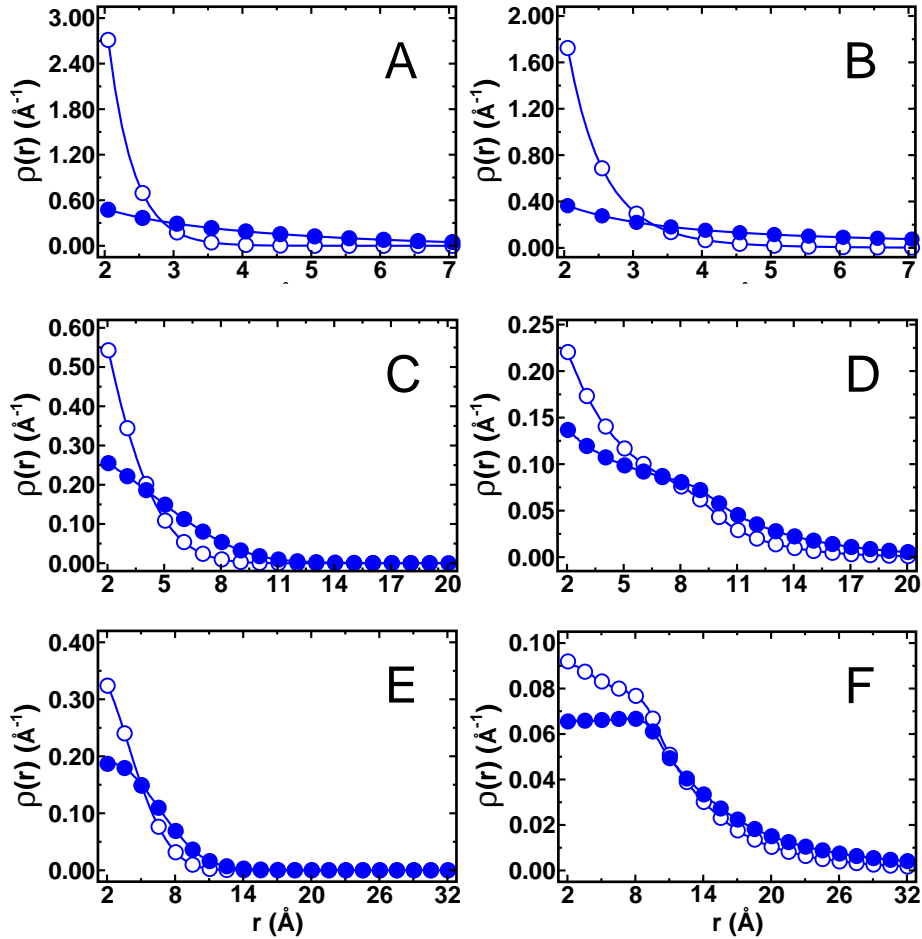

\includegraphics[width=6cm]{fig-dist-plna.eps}
\includegraphics[width=6cm]{fig-dist-plnb.eps}
\includegraphics[width=6cm]{fig-dist-plnc.eps}
\includegraphics[width=6cm]{fig-dist-plnd.eps}
\includegraphics[width=6cm]{fig-dist-plne.eps}
\includegraphics[width=6cm]{fig-dist-plnf.eps}
\caption{PE distributions near charged planar surfaces. The results obtained with the ES potentials from the linear and nonlinear PB equation are shown by open and closed symbols, respectively. Parameters: the surface charge density is $\sigma = -0.5$ (top panels), $-0.1$ (middle panels), and $-0.06$C/m$^2$ (bottom panels). The solution salinity is $n_0=10^{-3}$M (left panels) and 0.3M (right panels). The polymer chain contains $N=50$ monomers of  radius $R=2$\AA.}\label{fig-dist}\end{figure*}

At this point, we refer the reader to the study \cite{belt91} as to probably the first consideration of PE adsorption onto planar interfaces with the \textit{nonlinear} ES potential. Although a detailed consideration of the fraction of adsorbed monomers as a function of PE ionisation degree as well as the chain end-to-end extension was presented in this study, the explicit question of \textit{critical adsorption }conditions for the nonlinear ES potential has \textit{not }been addressed. That is why we here exploit this novel element of PE adsorption onto highly charged planar and convex interfaces with the nonlinear ES\ potentials.

Let us comment here on relatively high surface charge densities occurring in the text below. For comparison, the surface charge density of bare phosphate groups on the double stranded DNA is $\sigma_\text{DNA} \sim $ -0.16C/m$^2$ \cite{pars00, korn07, cher11b}.  Very high charge densitites are realised e.g. for cement paste  particles \cite{labb05}; their self-assembly with linear block copolymers in the presence of divalent Ca$^{2+}$ cations was examined in Ref. \cite{labb14}. The charge density of computer simulated cement platelets of up to $\sim$-0.64C/m$^2$ agrees with the estimations for C-S-H particles,  $\sim$-0.8C/m$^2$ \cite{labb05,labb14}. Other examples of highly charged interfaces/particles used for PE adsorption---such as i.a. silica, mica, and polysterene latex particles---can also reach $|\sigma|\sim$0.1-0.5C/m$^2$ \cite{blaa90, atal14, roja02, zsom87, anti98}. For some proteins, the patches of charges on their surfaces reveal large variations, from the charge densities of $\sim|\sigma_\text{DNA}|$ to considerably larger values in some cases \cite{barl86}.

\section{Results}
\label{sec-res}

\subsection{Adsorption onto Planar Surfaces}
\label{sec-planar}

We start with evaluating the adsorption of a single PE chain onto a planar surface with varying charge density $\sigma$. In Fig. \ref{fig-dist} we show how the PE monomer density profile $\rho(r)$ near the interface changes for PE adsorption driven by the nonlinearly (Eq. (\ref{eq-plane-nlpb})) versus linearly treated (Eq. (\ref{lpb-plan})) ES potential. We find that, in general, \textit{smaller} monomer concentrations are systematically observed in the immediate vicinity of the interface with the nonlinear ES potential. In this case, the PE segments rebind from the interface more readily forming some "dangling tails", whereas for the linear potential the chains are attached stronger to  the surface, often only diffusing along the interface while being fully adsorbed, see the video files in the Supplementary Material. The difference between the two approaches with regard to $\rho(r)$ decreases with increasing salinity, as expected, compare the left and right panels in Fig. \ref{fig-dist}. Considering the polymer distributions obtained with the nonlinearly treated potential at 0.3M of salt, the decrease of the surface charge density from $\sigma = -0.5$ (Fig. \ref{fig-dist}B) to $-$0.1C/m$^2$ (Fig. \ref{fig-dist}D) leads to the emergence of a region near the surface ($r < 10$\AA) with a slower decay of $\rho(r)$. Reducing the surface charge density even further, down to $-0.06$C/m$^2$ as in Fig. \ref{fig-dist}F, the polymer attains a state close to its desorption threshold at this solution salinity. The monomers assume a nearly constant concentration in the proximity of the surface, suggesting that the PE near the interface has a structure with various looped polymer segments, as detected in simulations. 

\begin{figure}\includegraphics[width=6cm]{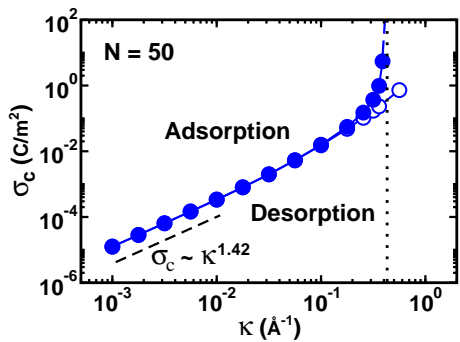}
\caption{Critical  charge density $\sigma_c$ for PE adsorption onto oppositely charged plane as a function of the Debye parameter $\kappa$. The polymer has $N=50$ monomers with   $R=2$\AA. The results for ES potentials from the linear ($\sigma_c^{l}$) and nonlinear ($\sigma_c^{nl}$) PB equations are the open and filled symbols, respectively. The maximal salinity at which PE adsorption still occurs is indicated by the vertical line. The error bars are smaller than the symbol size, but increasing as the adsorption--desorption boundary is approached, as expected.} \label{fig-sig1}\end{figure} 

In Fig. \ref{fig-sig1} we present the dependence of the critical adsorption charge density $\sigma_c$ on the solution salinity. As stated in Sec. \ref{sec-mod}, for a given salinity the value $\sigma_c$ defines the surface charge density at which we detect the same number of configurations in the adsorbed and desorbed states in the course of simulations. This is achieved via performing computations at varying charge density to meet this condition on average (at a given salinity and chain length $L=bN$). As we indicate in Fig. \ref{fig-sig1}, for the surface charge densities above the transition line the PE chains are in the adsorbed state due to strong ES\ attraction to the surface, while below this line the entropic free energy of the polymer dominates and the chain assumes on average desorbed configurations more often. For low salt concentrations we get a close agreement between the results obtained with the linear and nonlinear ES potentials. In this ionic strength regime, the critical charge density is low enough so that the corresponding ES potentials $\Psi(x)$ are close. 

\begin{figure}
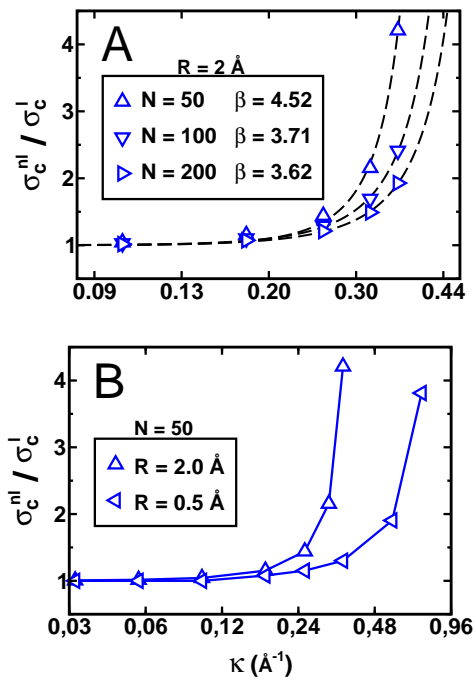

\includegraphics[width=6cm]{fig-div-plna-exp2.eps}
\includegraphics[width=6.2cm]{fig-div-plnb.eps}
\caption{Ratio of critical surface charge densities obtained with the linear and nonlinear ES potentials for the planar surface, plotted as a function of the Debye screening parameter $\kappa$. In panel A, the monomer radius is $R=2$\AA~and the chain length is $N$=50, 100, and 200. In panel B, the chain length is $N=50$ and the monomer radii are $R=0.5$ and 2\AA. The asymptotes of Eq. (\ref{eq-ratio-nonlinear-linear}) are shown as the dashed lines for the $\beta$ values as indicated in the legend of panel A.}
\label{fig-sigNR} \end{figure}

The cubic scaling of Eq. (\ref{eq-sigma-3}) is  not recovered here however. Instead, we get a weaker dependence of $\sigma_c$ on the solution salinity, namely \begin{equation}\sigma_c(\kappa) \sim \kappa^{\nu} \label{eq-sigma-nu} \end{equation} with the exponent $\nu\approx 1.42$ in the low salt limit. As discussed in Refs. \cite{cher14,muth87,kong98}, this change of the scaling exponent can be attributed to a $\kappa$-dependent ES contribution to the polymer persistence length \cite{muth94, wink14, muth87, cher14} as well as to some finite length effects. These features are taken into account in computer simulations---contrary to the theoretical models use to study critical PE-surface adsorption---giving rise to a quite different scaling exponent than the idealised cubic dependence  (\ref{eq-sigma-3}). 

In the limit of high salt we find that for nonlinearly treated ES potential the desorption transition occurs at \textit{substantially higher} surface charge densities, in comparison to the linear PB approach. Due to the nonlinear $\Psi(\sigma)$ coupling, the growth of the surface potential with the increasing charge density becomes weaker, see Fig. \ref{fig-pot}. So, the effective density of charges on the nonlinearly treated interface should be \textit{larger }to reach the same impact on the PE chain in front of it and ultimately to cause its electrostatically driven adsorption.  

Furthermore, this fast-growing value of $\sigma_c$ at high solution salinities appears to impose a physical limit onto the ionic strength beyond which the adsorption is \textit{fully suppressed}. This effect is not observed for the linearly treated ES potential because of the linear relation between the surface potential and the charge density, see Eq. (\ref{eq-psi-sigma}). This made it possible to increase the surface charge density and always  have sufficiently large  ES potentials to stabilise the PE adsorption. 

The difference of the results obtained with the two approaches can be gauged from Fig. \ref{fig-sigNR}. It illustrates the ratio between the critical surface charge densities obtained with the linear ($\sigma_c^{l}$)
and nonlinear ($\sigma_c^{nl}$) PB equations, both as the functions of  $\kappa$. In panel A, we consider three 
different degrees of chain polymerisation. We find that the difference between $\sigma_c^{nl}$ and $\sigma_c^{l}$ becomes considerable for $\kappa$ values larger than $\sim1/(5$\AA) for all chain lengths examined. However, the deviations are  highly dependent on the polymer length. Specifically, we find that longer chains---for which the critical charge density is smaller in magnitude---reveal smaller deviations in $\sigma_c$, as expected. The deviations between the two ES approaches start occurring at about the same ionic strength as the polymer gets "flattened" onto the surface, as can be seen in Fig. \ref{fig-intro} plotted for $\kappa\approx0.27$/\AA~ and from the video files in the Supplementary Material. 

One feature of the full nonlinear ES potential is the fact that its variation with increasing surface charge density becomes larger in closer proximity to the surface, see the top panel in Fig. \ref{fig-pot}. This has vital implications for the PE adsorption properties, as we show in Fig. \ref{fig-sig1} and below. We find, for instance, that when the thickness of the interfacial layer with relatively high values of the ES potential becomes considerably smaller than the monomer diameter $2R$, the changes of the surface charge density only \textit{weakly affect }the attraction of the closest PE monomers. This can be seen in Fig. \ref{fig-sigNR}B where we plot the ratio $\sigma_c^{nl}/\sigma_c^{l}$ for critical adsorption of PEs, computed for two different monomer diameters. Indeed, smaller monomers can approach closer to the attracting surface, into the regions of higher ES\ potentials. Thus, the deviations between $\sigma_c^{nl}$ and $\sigma_c^{l}$ start occurring at higher salinities for the chains with smaller monomers. 

We can approximate the behaviour of the critical charge density with solution salinity by a functional dependence of the form \begin{equation} \sigma_c^{nl}(\kappa)/ \sigma_c^{l}(\kappa)\sim\exp[\kappa^\beta]. \label{eq-ratio-nonlinear-linear} \end{equation} Here, the exponent $\beta$ however depends on the monomer radius and chain length. The simulations data presented in Fig. \ref{fig-sigNR} support the phenomenological dependence (\ref{eq-ratio-nonlinear-linear}). This exponential variation should emphasise how unstable the PE adsorption becomes in the realistic nonlinear PB theory for the conditions of large salinities, when the region of a high ES potential near the interface becomes progressively thinner. Also, we find that at a given and relatively large  $\kappa$ for longer chains the deviations of linear versus nonlinear critical adsorption surface charge densities are smaller than for shorter chains, see Fig. \ref{fig-sigNR}A. One of the reasons is that for longer chains a substantial PE portion far from the interface is still in the region of ES potential with a nearly exponential decay, similar to the linear PB solution. Finally, the PE chains with smaller monomers experience the deviations in $\sigma_c$ for the full nonlinear versus the linearised ES potential at larger salinities, as intuitively expected, see Fig. \ref{fig-sigNR}B.

\subsection{Adsorption onto Curved Surfaces}
\label{sec-curved}

\begin{figure}\includegraphics[width=6cm]{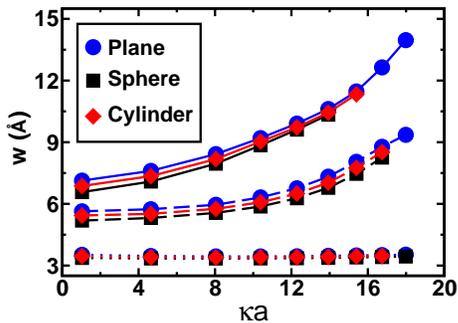}
\caption{Width  $w$ of the adsorbed PE  layer onto the planar (blue circles), cylindrical (red diamonds) and spherical (black squares) surfaces as a function of $\kappa a$, obtained with computer simulations for the nonlinear ES potential. The charge density is $\sigma = 0.06$ (solid line), $0.1$ (dashed line) and $0.5$ C/m$^2$ (dotted line). The cylinder and sphere radius is $a=100$\AA~and the salinity is varied. The chain contains $N$=50 beads of radius $R=2$\AA.}\label{fig-tick}\end{figure}

In this section, we address the properties of PE adsorption onto cylindrical and spherical interfaces with the curvature radius $a=100$\AA. In Fig. \ref{fig-tick} we present the results of computer simulations for the width of the adsorbed PE layer  $w$ obtained with the nonlinearly treated ES potential, see Eqs. (\ref{eq-nlpb-sphere-final-uni-solution}) and (\ref{eq-solution-uni-cyl}) for the spherical and cylindrical surfaces, respectively. The layer thickness  $w$ and the amount of adsorbed polymer are the PE adsorption observables accessible experimentally via e.g. the ellipsometry and total light reflection measurements \cite{kawa82}. Similarly to our previous studies \cite{cher11,wink14}, the width of the adsorbed PE layer is computed at the half height of the polymer monomer density profile, $\rho(r)$. Here and below the results of the PE\ adsorption for the linearised ES\ potential are shown as the open symbols, while the findings for the nonlinear potential are the filled symbols. The results for the three standard geometries are designated by the blue (plane), red (cylinder), and black (sphere) symbols.

As expected, we observe a systematic increase of the PE layer thickness $w$ with the ionic strength. This stems from the fact that at higher salinities more monomers stay in "looped" and "tailed" conformations, see also Refs. \cite{belt91,kawa82}, not necessarily very close to the attracting surface. The PE layer thickness is shown in Fig. \ref{fig-tick} for the range of ionic strengths when the PE\ adsorption is stable and for   $\sigma$ values well above the adsorption--desorption transition. In all three geometries, we find that with \textit{increasing } $\sigma$ the fraction of the chains in tailed and looped states decreases, while the fraction of tightly adsorbed monomers in "train-like" conformations \cite{kawa82,golu01} gets larger. The same trends are observe when the screening length decreases, thus facilitating the ES attraction of PEs to the interface. These pieces of evidence are consistent with the properties of PE adsorption presented in Ref. \cite{belt91}  based on the full nonlinear ES potential. 

\begin{figure}\includegraphics[width=6cm]{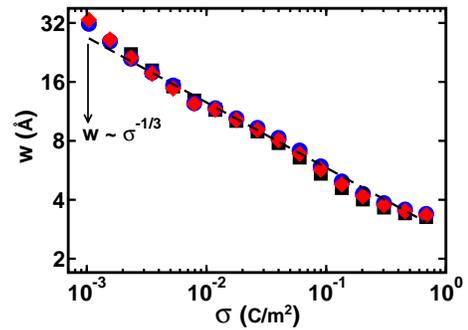}
\caption{Width of the adsorbed PE layer versus the surface charge density $\sigma$ for the three adsorption geometries. The asymptote of Eq. (\ref{eq-w}) is the dashed line. The notations for the symbols are the same as in Fig. \ref{fig-tick}. Here, the chain of  $N=$50 beads with monomer radius  $R=2$\AA\ is immersed into the solution with   $\lambda_D=100$\AA. The radius of curvature of spherical and cylindrical surfaces is $a$=100\AA.}\label{fig-tick-2}\end{figure}

We also note that with increasing $\kappa$ the polymer-surface ES interactions decrease, as compared to the entropic free energy penalty in the course of polymer confinement near the interface. The latter depends on the surface geometry  \cite{cher11} so that for a relatively small surface curvature, the changes in the surface geometry lead to only slight variations of the layer width, compare the curves for different geometries in Fig. \ref{fig-tick}.

\begin{figure*}
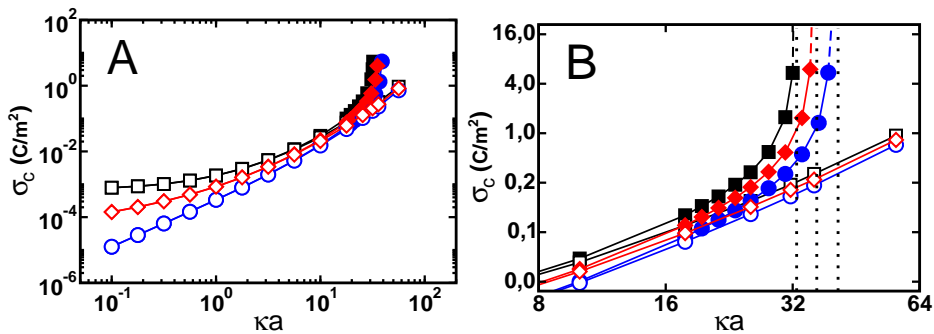

\includegraphics[width=6cm]{fig-crit-geoa.eps}
\includegraphics[width=6.2cm]{fig-crit-geob.eps}
\caption{Critical surface charge density for PE adsorption onto the planar (blue circles), cylindrical (red diamonds) and spherical (black squares) surfaces as a function of $\kappa a$. The notations are the same as in Fig. \ref{fig-tick}. Panel B magnifies the region in which the rapid changes in $\sigma_c$ occur, with vertical dotted lines indicating the maximal salinity still enabling PE adsorption. The cylinder and sphere radius is $a=100$\AA~and the salt concentration  is varied. The polymer contains $N$=50 beads of radius $R=2$\AA. The results obtained with the linearly and nonlinearly treated ES potential are shown by open and closed symbols, respectively. On a standard 3-3.5 GHz workstation every curve on these graphs requires  $\sim$150 hours of computation time.} \label{fig-crit-geo}\end{figure*} 

Let us now consider the variation of the PE\ layer thickness with the surface charge density $\sigma$. Some analytical predictions exist regarding $w(\sigma)$ dependencies, namely, the scaling behaviour \begin{equation} w(\sigma)\sim\sigma^{-1/3}. \label{eq-w} \end{equation} is often advocated, see e.g. Refs. \cite{dobr05,cher11,wink14}. The reader is also referred here to our recent study \cite{cher14} regarding the thickness of the adsorbed PE layer near the  dipolar Janus particles. The results for the PE\ layer thickness for the nonlinear ES potential are presented in Fig. \ref{fig-tick-2}. They are consistent with the general trend of Eq. (\ref{eq-w}) for substantial surface charge densities. For progressively lower $\sigma$, however, the layer width grows and ultimately diverges at the \textit{critical point }at which the transition from adsorption to desorption takes place. At such conditions, however, the PE is either in the adsorbed or in the desorbed state, so a statistically meaningful determination of the average PE layer thickness is not possible. In Fig. \ref{fig-tick-2}, in the region of even smaller surface charge densities, the polymer chain assumes a desorbed state and the layer width $w$ becomes effectively infinite, see the last points computed in the region of small $\sigma$  in the corresponding geometry.  

The reader may compare compare our simulations results of Fig. \ref{fig-tick-2} to the theoretical results of the linear PB theory presented in Fig. 5 in Ref. \cite{cher11}. We also observe that---as compared to the planar surface with the same charge density---the width of the adsorbed layer is \textit{slightly larger }for the cylindrical and even larger for the spherical interface. This is in line with the physical intuition that polymer adsorption onto convex interfaces gives rise to stronger restriction on polymer conformations, to a higher entropic free energy penalty upon chain confinement, and thus leads to a weaker attachment of PE chains to the interface. The differences in $w$ become larger for strongly curved interfaces, as compared to rather large value  of surface curvature radius $a=100$\AA~ used in Fig. \ref{fig-tick-2} (results not shown).

Finally, we investigate the adsorption--desorption conditions in the three standard adsorption geometries. Fig. \ref{fig-crit-geo} illustrates that for low-to-moderate salinities, for these curved surfaces we find \textit{excellent agreement} between the results obtained with the linearly and nonlinearly treated ES potentials (details are given in App. \ref{app-potentials}). This is similar to the results for the planar interface in Fig. \ref{fig-sig1}. The scaling exponent $\nu$ for the  $\sigma_c(\kappa)$ relation  (\ref{eq-sigma-nu}) in the \textit{low salt }regime decreases systematically for the cylindrical and spherical surfaces, in comparison to the plane. This in
agreement with the analytical results of the ground state approximation for the linear PB equation derived in Ref. \cite{cher11}. At \textit{high }ionic strengths, in contrast, we observe for the spheres and cylinders a rapid increase of the critical surface density, qualitatively similar to that observed for the planar surface in Fig. \ref{fig-sig1}. 

The curvature effects at \textit{high salt }can be attributed to the salinity at which the abrupt growth of $\sigma_c(\kappa)$ emerges. For the case of linear ES potential, in this regime we observe rather \textit{small }variations of  $ \sigma_c$ obtained for different geometries, as shown in Fig. \ref{fig-crit-geo}. These are due to curvature effects and larger confinement penalty of the polymer  near the attracting \textit{convex }surfaces. For the nonlinearly treated ES potential the geometry mediated deviations in $ \sigma_c$ become substantially larger. In particular, the simulations reveal that spherical interfaces cease to adsorb PE chains at lower salinity, as compared to the cylindrical interfaces. The same is true when comparing the cylindrical and planar interfaces, see the corresponding dotted lines in Fig. \ref{fig-crit-geo}B indicating the limiting salinity values for each adsorption geometry. 

\section{Discussion and Conclusions}
\label{sec-dis}

We carried out extensive computer simulations to unravel the properties of electrostatically driven adsorption of flexible PE chains onto oppositely charged surfaces of arbitrarily high surface charge densities. We used the known exact solution for the screened ES potential of the plane and approximate solutions of the full nonlinear PB equation in spherical and cylindrical geometries \cite{ohsh82}. Our findings revealed  a number of important new features, as compared to the PE adsorption properties expected from the linear PB theory. 

In particular, we demonstrated that a nonlinear dependence of the surface ES potential and the surface charge density lead to an \textit{abrupt increase} of the critical adsorption charge density at high salinities of the solution. In the region prior to and at this increase of the critical adsorption charge density, \textit{no distinct scaling} with the solution salinity is detected in simulations, see Fig. \ref{fig-sig1}. At low salinities, in contrast, the results for the critical adsorption charge density obtained with the linear and nonlinear ES potential treatments superimpose, as they should. In this limit, the surface charge density required for the PE adsorption is rather small, so the linear PB theory is valid for ES\ potential calculations. The full nonlinear ES potential of the adsorbing surface also imposes a limit onto the ionic strength above which \textit{no adsorption takes place at all}. This limiting ionic strength depends on the  surface geometry, being smaller for the spherical surface, intermediate for the cylindrical interface, and maximal for the planar adsorbing boundary, see Fig. \ref{fig-crit-geo}B. In addition, we described the conformations of \textit{partly adsorbed }PE chains in terms of the width of the polymer layer $w$ near the interface. The results obtained for the $w(\sigma)$ dependence from the full nonlinear PB theory are in good agreement with the general theoretical predictions, for all adsorption geometries studied.

One immediate application of our results is related to the PE multilayer formation \cite{dech97,cher14c,pogh13,scho03,scho07,volo14}, governed by ES driven complexation of alternating oppositely charged PE chains and the release of water molecules forming hydration shells around them, see Refs. \cite{borg14,peyr04,hamm99, dobr05,lasc00} for the details of physical complexation mechanisms. For the PE chains adsorbed in such multilayers, a systematic and well controlled layer growth of  oppositely charged polymers is typically achieved at low salt concentrations in the bulk. At these conditions, rather thin and compact PE layers are formed on the substrate. For the classical example of PAH-PSS multilayers, the magnitude of voltage variations on the outer PE layer in the course of multilayer formation can reach $\sim$100mV at 1mM of simple salt \cite{pogh13, cher14c}. This is clearly beyond the applicability regimes of the linear PB theory and the full nonlinear ES potential calculations need to be used. 

Our results can also help understanding the features of macromolecular adsorption onto highly charged surfaces as found, for instance, on silica nanoparticles. The latter are widely used for the multilayered PE deposition \cite{dech97,dech12, fors12,imot10}, with the surface charge magnitudes reaching $\gtrsim$0.1C/m$^2$   \cite{schu97, abba16, bari14}. The $\zeta$-potential of silica particles coated by alternating deposition of chitosan and poly-($\gamma$-glutamic acid) PEs was demonstrated to vary between +60 and -40mV. These values are again beyond the reach of the linear Debye-H\"uckel theory. The nonlinear approach is to be implemented for the quantitative description of potential variations. Similar values of the ES potential variations were observed when using the liposomes as a template to build some PE nanocapsules \cite{lind12}. Note also the properties of electrostatically driven adsorption of proteins onto highly charged silica surfaces and inside porous substrates, see Refs. \cite{hube14, hube15a, hube15b}. 

At all these conditions, the nonlinear ES effects can come into play. Thus, close to the adsorption--desorption transition rather small variations of the ionic strength or pH can may cause significative changes on PE conformations and alter the basic adsorption characteristics of PE chains, as we demonstrated in this study.

\section{Acknowledgements}

Computer resources were supplied by the Center for Scientific Computing (NCC/GridUNESP) of the Sao Paulo State University. The authors acknowledge funding from the Funda\c c\~ao de Amparo \`a Pesquisa do Estado de S\~ao Paulo (FAPESP Proc. 15/15297-4 to SJC).


\numberwithin{figure}{section}
\renewcommand{\thefigure}{A\arabic{figure}}

\appendix{} \section{} \label{app-potentials}

Here, we present the approximate solutions of the nonlinear PB equation for the distribution of the ES potential near spherical and cylindrical interfaces. We closely follow the procedure developed in Refs. \cite{ohsh82,ohsh98} for obtaining the uniformly valid solutions for the potential and the generalised Grahame relations on the interfaces. These approximate potential distributions have the same structure as the one near a planar surface and yield the correct limiting behaviours both in the close field and in the far field regions. The corresponding Grahame relations for the curved highly charged interfaces are obtained \cite{ohsh82} via integrating the PB equation and using the boundary conditions for the ES potential and its derivative. These relations provide the nonlinear coupling of the surface charge density and surface potential, in contrast to the linear $\Psi_s(\sigma)$ relation in the Debye--H\"uckel theory.

\subsection{Planar Geometry} 

We start with the solution for the dimensionless ES potential $$\Psi(\textbf{r})=e_ 0\phi (\textbf{r})/(k_B T)$$ emerging near a  planar surface immersed in 1:1 electrolyte solution with the salt concentration $n_0$. This solution for the nonlinear PB equation \begin{equation} d^2\Psi(x)/dx^2=\kappa^2\sinh(\Psi(x)) \label{eq-nlpb-eq}\end {equation} at a constant surface potential $y_s$ is the well known expression from the  Gouy--Chapman theory of electrical double layer formation, namely \cite{ande95, kapp06} \begin {equation} \Psi(x) =  2\log\left[\frac {1 + \tanh (\Psi_s/4) e^{-\kappa x}}{1 - \tanh (\Psi_s/4) e^{-\kappa x}}\right]. \label{eq-plane-nlpb} \end{equation} The standard boundary conditions are that the potential at the charged surface assumes the value $\Psi_s$ and far away from the charged interface the ES potential and its derivative both vanish, $\Psi(x) = d\Psi(x)/dx = 0$ at $x\to\infty$. For the surface charge density being represented as \begin{equation}\sigma = e_ 0/S,\label{eq-sigma} \end{equation} where $S$ is the surface per elementary charge $e_ 0 $, the Grahame relation \cite{ande95,kapp06} from the nonlinear PB equation is \begin {equation} 4\pi l_B/(S\kappa) = 2\sinh (\Psi_s/2). \label{eq-gram-plane} \end{equation} Naturally, in the limit of small potentials and weakly charged surfaces, when the linear PB theory of diffuse double layers is valid and \begin{equation} \Psi (x) = \Psi_s e^{-\kappa x},\label{lpb-plan}\end{equation} this relation turns into the standard \begin {equation} d\phi(x)/dx|_{x=0} = -4 \pi \sigma/\epsilon. \label{eq-linear-psi-sigma-relation} \end {equation} 

\subsection{Spherical Geometry} 

For the spherical geometry in terms of the dimensionless variable \cite{ohsh82} \begin{equation}R = \kappa r\end{equation} the nonlinear PB equation to be solved becomes \begin {equation} d^2 \Psi (R)/dR^2 + (2/R) d\Psi (R)/dR = \sinh (\Psi (R)). \label{eq-nlpb-eq-sphere}\end {equation} Provided the same standard boundary conditions are imposed, introducing new dimensionless variables \begin{equation}\xi =\kappa(r-a)\label{eq-xi-via-r}\end{equation} and \begin{equation}A=\kappa a,\label{eq-a-defin}\end{equation} we rewrite Eq. (\ref{eq-nlpb-eq-sphere}) as \begin {equation} \frac{d^2 \Psi(\xi)}{d\xi^2} = \sinh (\Psi (\xi)) - \frac {2} {A}\frac {A} {A + \xi} \frac{d\Psi (\xi)}{d\xi}.\label{eq-nlpb-eq-sphere-app1}\end {equation} Using the result of the first integration of the planar Eq. (\ref{eq-plane-nlpb}) as the initial approximation, namely that $d\Psi(\xi)/d\xi = -2\sinh (\Psi(\xi)/2),$ we arrive in the limit $A\gg 1 $ at \begin {equation} d^2\Psi(\xi)/d\xi^2 = \sinh (\Psi (\xi)) + (4/A)\sinh (\Psi(\xi)/2).\label{eq-nlpb-eq-sphere-app2}\end {equation} The integration of this equation gives \begin {equation} \frac{d\Psi (\xi)}{d\xi} = 2\sinh \left(\frac{\Psi (\xi)}{2}\right)\sqrt {1 + \frac {2} {A}\frac {1} {\cosh^2 (\Psi (\xi)/4)}}.\label{eq-first-der-sphere}\end {equation} At the interface at $\xi = 0 $ this yields the next order approximation for the Grahame relation for the nonlinear PB in the spherical geometry. Namely for the dimensionless surface charge density defined as $$I (\Psi_s) = -d\Psi(\xi)/d\xi|_{\xi=0} ={4 \pi e_0 \sigma (\Psi_s)}/({\epsilon \kappa k_B T})$$ we get \begin {equation} I (\Psi_s) = 2\sinh (\Psi_s/2) + (4/A)\tanh (\Psi_s/4).\label{eq-grahame-sphere-2}\end {equation} This relation in the limit of small potentials gives \begin {equation} I (\Psi_s) = \Psi_s (A + 1)/A, \label{eq-I-lpb-theory-limit}\end {equation} that is the correct result following from the exact ES potential solution for the sphere \cite{wink14}, namely \begin {equation} \Psi(r) = \frac {4 \pi \sigma a^2 e^{-\kappa (r-a)}} {\epsilon (1 + \kappa a) r}\frac {e_ 0} {k_BT}. \label{eq-lpb-solution-potential} \end {equation}  

\begin{figure}\includegraphics[width=8cm]{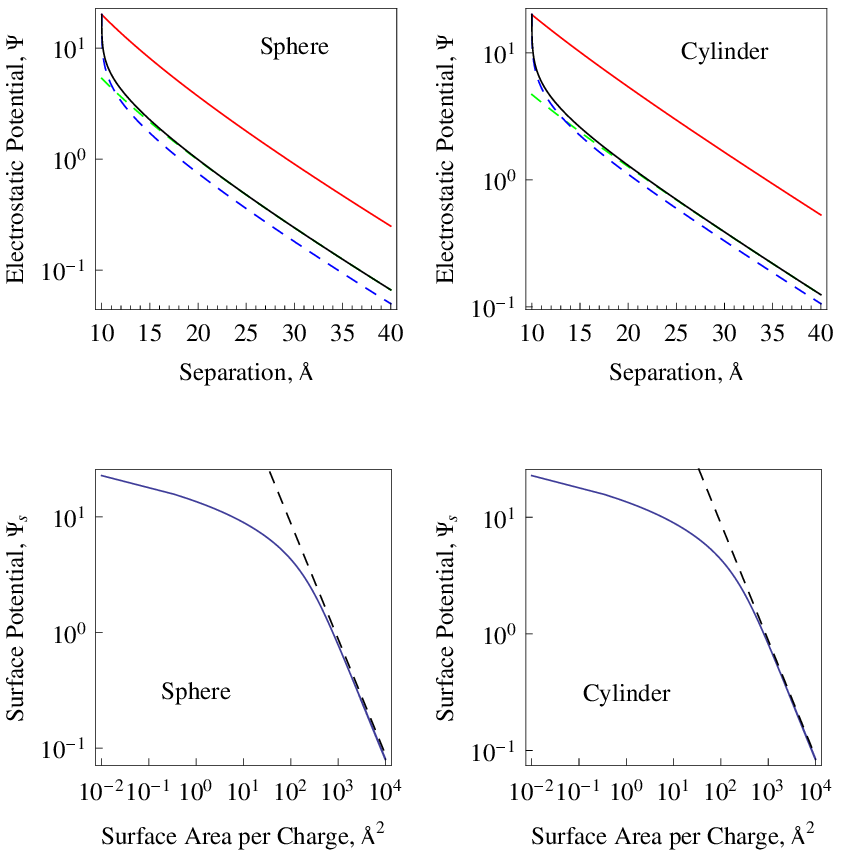}
\caption{Decay of the linear and nonlinear PB potential in spherical and cylindrical geometries (top panels) and the nonlinear relation between the surface charge density and the surface potential (bottom panels). In the top panel, the linear PB potential is shown by the red curve, while the full nonlinear PB solution is the solid black curve. The close field and far field asymptotes for the ES potential given by Eqs. (\ref{eq-far-field-sphere}) \& (\ref{eq-pot-sphere-small-s-limit}) and Eqs. (\ref{eq-far-field-cyl}) \& (\ref{eq-potential-approx-far-field}) for spherical and cylindrical geometries, respectively, are  the blue dashed and green dashed curves. The linear relation  (\ref{eq-linear-psi-sigma-relation}) for $\Psi_s (\sigma) $ is the dashed line in the bottom panel. Parameters: $a=100$\AA\ and  $\kappa=1/$(10\AA).}
\label{fig-pot}\end{figure}

After getting the approximate close field solution for the ES potential, we now find an approximate solution in the far field limit. For this, a new variable $s(r)$ is introduced to reflect the features of the far field behaviour of the potential, as known from the linear PB theory (\ref{eq-lpb-solution-potential}). Specifically, we use \begin {equation} s (r) = Ae^{-(R(r) - A)}/R(r) \sim e^{-\kappa (r - a)}/r.\label{eq-s-via-r}\end{equation} In terms of this variable Eq. (\ref{eq-nlpb-eq-sphere}) turns into \begin {equation} s^2 \frac{d^2 \Psi (s)}{ds^2} + s\frac{d\Psi (s)}{ds} = \sinh (\Psi (s)) - \frac {2 A + 1} {(A + 1)^2} G (\Psi (s)),\label{eq-nlpb-eq-sphere-app3}\end{equation}  where the function $G (y) $ is defined as \begin {equation} G (\Psi) = \frac {2 R + 1} {2 A + 1}\frac {(A + 1)^2} {(R + 1)^2} \left(\sinh (\Psi)-s\frac{d\Psi}{ds}\right). \label{eq-G-of-y}\end{equation} In the limit of weakly curved interfaces, i.e. when $A \gg 1 $ and in the range $R\sim A$, the last term in Eq. (\ref{eq-nlpb-eq-sphere-app3}) disappears. With the additional substitution of variables $s = e^t$  its solution satisfying the boundary conditions has the form of Eq. (\ref{eq-plane-nlpb}) for the planar surface, namely \begin {equation} \Psi (s) =  2\log\left[\frac{1 + \tanh (\Psi_s/4) s} {1 - \tanh (\Psi_s/4) s} \right].\label{eq-far-field-sphere}\end {equation} Here the fundamental solution of the linear PB equation in the spherical geometry, $e^{-\kappa (r - a)},$ plays the role of the decaying exponent $e^{-\kappa x} $ in the planar solution of Eq. (\ref{eq-plane-nlpb}), see Fig. \ref{fig-pot}. 

In the next order perturbation---similar to the procedure of the close field solution of Eq. (\ref{eq-nlpb-eq-sphere-app2})---we use the solution (\ref{eq-far-field-sphere}) to get $G (\Psi)\approx \sinh (\Psi) -  2\sinh (\Psi/2) $. Then Eq. (\ref{eq-nlpb-eq-sphere-app3}) with the substitution $s = e^t$ can be integrated once to give \begin{equation} \frac{d\Psi}{dt} = s\frac{d\Psi}{ds} = \frac {2A} {A + 1} \sinh \left(\frac{\Psi}{2}\right) \sqrt {1 + \frac {2 A + 1} {A^2}\frac {1} {\cosh^2 (\Psi/4)}}. \label{eq-sphere-first-der-far-field} \end {equation} From this relation the generalised Grahame equation is $I = s(d\Psi/ds)|_{s=1} (A + 1)/A$ or in terms of the surface area per charge \begin {equation} \frac{4\pi l_B}{S\kappa}= 2\sinh \left(\frac{\Psi_s}{2}\right) \sqrt {1 + \frac {2 A+ 1} {A^2}\frac {1} {\cosh^2 (\Psi_s/4)}}.\label{eq-grahame-sphere-3} \end{equation} This relation yields the correct limits of Eq. (\ref{eq-grahame-sphere-2}) for a small curvature $A \gg  1 $ and of Eq. (\ref{eq-I-lpb-theory-limit}) for small ES potentials $\Psi_s \ll 1 $. Integrating Eq. (\ref{eq-sphere-first-der-far-field})  we get the \textit{uniformly valid} solution of the nonlinear PB equation in the spherical geometry, \begin {equation} \Psi (s) = 2\log\left[\frac {1 + Bs} {1 - Bs}\frac {1 + \frac {B} {2 A + 1}s} {1 - \frac {B} {2 A + 1} s} \right],\label{eq-nlpb-sphere-final-uni-solution}\end{equation} where the function $B$ depends on the surface potential and has the form \begin {equation} B=B (\Psi_s) = \frac {\tanh (\Psi_s/4) (1 + A/(A + 1))} {1 + \sqrt {1 - \frac {2 A + 1} {(A + 1)^2}\tanh^2 \left(\frac{\Psi_s}{4}\right)}}. \label{eq-B-of-ys-function}\end{equation} In the limit of weak potentials the general expression (\ref{eq-nlpb-sphere-final-uni-solution}) gives $\Psi (s)\approx \Psi_s s$, as expected. Also, in the far field region---when $R\to\infty$ and $s\to 0 $---this expression in the leading order gives the potential variation linear in variable $s$, namely \begin{equation} \Psi (s) \approx  \frac {8\tanh (\Psi_s/4) s} {1 + \sqrt {1 - \frac {2 A + 1} {(A + 1)^2}\tanh^2 \left(\frac{\Psi_s}{4}\right)}}.\label{eq-pot-sphere-small-s-limit}\end {equation}

\subsection{Cylindrical Geometry} 

Likewise, for the ES potential in the nonlinear PB theory in the cylindrical geometry, \begin {equation} d^2 \Psi (R)/dR^2 + (1/R) d\Psi (R)/dR = \sinh (\Psi (R)),\label{eq-nlpb-eq-cyl}\end {equation} we introduce a new variable $c$ \cite{ohsh82}. It reflects the far field behaviour of the linear ES potential, namely \begin {equation} c (r) = K_ 0 (\kappa r)/K_ 0 (\kappa a).\label{eq-c-var}\end {equation} Here $K_0(z)$ is the modified Bessel function of the second kind. It describes the potential decay from a uniformly charged cylinder in the linear PB scenario \cite{wink14}, \begin {equation} \Psi(r) = \frac{4 \pi \sigma a}{\epsilon} \frac{K_ 0 (\kappa r)}{\kappa a K_1(\kappa a)}\frac{e_0}{k_BT}. \label{eq-sol-lpb-cyl}\end {equation} Eq. (\ref{eq-nlpb-eq-cyl}) can then be presented in the form \begin {equation} c^2 \frac{d^2 \Psi (c)}{dc^2} + c\frac{d\Psi (c)}{dc} = \sinh (\Psi (c)) - \left(1 - \frac {K_ 0^2 (A)} {K_ 1^2 (A)}\right) H (\Psi (c)),\label{eq-nlpb-cyl-app1}\end {equation} where the function $H (\Psi) $ is \begin {equation} H (\Psi) = \frac {1 - \frac {K_ 0^2 (R)} {K_1^2 (R)}} {1 - \frac {K_ 0^2 (A)} {K_1^2 (A)}} \left(\sinh (\Psi) - c\frac{d\Psi}{dc}\right). \label{eq-H-of-c-function}\end {equation} In the limit of small curvature $A \gg 1 $ the dependence on $A$ disappears on the right hand side of Eq. (\ref{eq-nlpb-cyl-app1}), so that one gets $c^2 d^2 \Psi/dc^2 + cd\Psi/dc \approx \sinh (\Psi).$ Its solution in the close field limit is similar to the planar nonlinear PB solution (\ref{eq-plane-nlpb}), namely (see Fig. \ref{fig-pot}) \begin {equation} \Psi (c) =  2\log\left[\frac {1 + \tanh (\Psi_s/4) c} {1 - \tanh (\Psi_s/4) c} \right]. \label{eq-far-field-cyl}\end {equation} Using the first derivative of this potential,  one arrives at $H (\Psi) \approx \sinh (\Psi) - 2 \sinh (\Psi/2) $ and \begin {eqnarray}\nonumber c^2 \frac{d^2 \Psi (c)}{dc^2} + c\frac{d\Psi (c)}{dc} = \sinh (\Psi(c)) -\\  \left(1 - \frac {K_ 0^2 (A)} {K_ 1^2 (A)}\right) \left[\sinh (\Psi (c)) - 2 \sinh \left(\frac{\Psi (c)}{2}\right)\right]. \label{eq-nlpb-cyl-app11} \end {eqnarray} The solution of this equation is \begin {equation} \Psi (c) = 2\log\left[\frac {1 + Dc} {1 - Dc}\frac {1 + \left({1 - \frac {K_ 0 (A)} {K_ 1 (A)}}\right) Dc\left/\left({1 + \frac {K_ 0 (A)} {K_ 1 (A)}}\right)\right.} {1 - \left({1 - \frac {K_ 0 (A)} {K_ 1 (A)}}\right) Dc\left/\left({1 + \frac {K_ 0 (A)} {K_ 1 (A)}}\right)\right.} \right],\label{eq-solution-uni-cyl}\end {equation} where the function $ D $ has the form \begin {equation} D=D (\Psi_s) = \frac {\tanh\left(\frac{\Psi_s}{4}\right)\left(1 + \frac {K_ 0 (A)} {K_1 (A)}\right)} {1 + \sqrt {1 - \left(1 - \frac {K_0^2 (A)} {K_ 1^2 (A)}\right) \tanh^2 \left(\frac{\Psi_s}{4}\right)}}.\label{eq-D-function}\end {equation} In the limit of small potentials Eq. (\ref{eq-solution-uni-cyl}) yields \begin{equation} \Psi(r)\approx \Psi_s c = \Psi_s K_ 0 (\kappa r)/K_ 0 (\kappa a) \end{equation} as it should, see Eq. (\ref{eq-sol-lpb-cyl}). 
The far field expansion of the ES potential at $c\to 0 $ has---again similarly to the spherical geometry---the decay reminiscent to that of the linear PB potential (\ref{eq-sol-lpb-cyl}). Namely, we get \begin {equation} \Psi (r) \approx \frac {8\tanh \left(\frac{\Psi_s}{4}\right)\frac {K_ 0 (\kappa r)} {K_ 0 (A)}} {1 + \sqrt {1 - \left(1 - \frac {K_ 0^2 (A)} {K_ 1^2 (A)}\right)\tanh^2 \left(\frac{\Psi_s}{4}\right)}}. \label{eq-potential-approx-far-field}\end {equation} The reader is also referred to the studies  \cite{lifs51, barr96} for the exact solution for the nonlinear ES potential distribution around a charged rod in a salt free regime (only the PE rod with own counterions) in the cylindrical Wigner--Seitz cell model. Finally, the generalised Grahame relation in the cylindrical geometry follows from the first integration of Eq. (\ref{eq-nlpb-cyl-app11}) using the condition of vanishing potential and its derivative at infinity, \begin {equation} I = \frac{4\pi l_B}{S\kappa}=2 \sinh \left(\frac{\Psi_s}{2}\right)\sqrt {1 + \frac {{K_ 1^2 (A)}/ {K_ 0^2 (A)} - 1} {\cosh^2 (\Psi_s/4)}}. \label{eq-grahame-cyl} \end {equation} In the limit of small potentials, the standard relation between the potential and surface charge density, \begin{equation}\phi_s(\sigma)= 4\pi\sigma/(\epsilon\kappa), \label{eq-psi-sigma} \end{equation} is naturally recovered from Eq. (\ref{eq-grahame-cyl}).






\begin{thebibliography}{99}

\bibitem{netz03} R. R. Netz and D. Andelman, 
\textit{Neutral and Charged Polymers at Interfaces}, 
Phys. Rep. \textbf{380}, 1 (2003).

\bibitem{dobr05}  A. V. Dobrynin and M. Rubinstein, 
\textit{Theory of Polyelectrolytes in Solutions and at Surfaces}, 
Prog. Polym. Sci. \textbf{30}, 1049 (2005).

\bibitem{dobr08} A. V. Dobrynin, 
\textit{Theory and Simulations of Charged Polymers: From Solution Properties to Polymeric Nanomaterials}, 
Curr. Opin. Colloid \& Interf. Sci. \textbf{13}, 376 (2008).

\bibitem{kawa82} A. Takuhashi and M. Kawaguchi, 
\textit{The Structure of Macromolecules Adsorbed on Interfaces}, 
Adv. Polym. Sci. \textbf{46}, 1 (1982).

\bibitem{wink14} R. G. Winkler and A. G. Cherstvy, 
\textit{Strong and Weak Polyelectrolyte Adsorption onto Oppositely Charged Curved Surfaces},
Adv. Polym. Sci. \textbf{255}, 1 (2014).

\bibitem{khok15} M. M. Feldstein, E. E. Dormidontova, and A. R. Khokhlov, \textit{Pressure Sensitive Adhesives based on Interpolymer Complexes},
Prog. Polym. Sci. \textbf{42}, 79 (2015). 

\bibitem{wieg77} F. W. Wiegel, 
\textit{Adsorption of a Macromolecule to a Charged Surface}, 
J. Phys. A: Math. Gen. \textbf{10}, 299 (1977).

\bibitem{wieg86} F. W. Wiegel, "Introduction to path-integral methods in physics and polymer science", (World Scientific, Singapore, 1986).

\bibitem{muth87} M. Muthukumar, 
\textit{Adsorption of a Polyelectrolyte Chain to a Charged Surface}, 
J. Chem. Phys. \textbf{86}, 7230 (1987).

\bibitem{joan91} D. Andelman and J.-F. Joanny, 
\textit{On the Adsorption of Polymer Solutions on Random Surfaces: the Annealed Case}, 
Macromolecules \textbf{24}, 6040 (1991)

\bibitem{muth94} F. von Goeler and M. Muthukumar, 
\textit{Adsorption of Polyelectrolytes onto Curved Surfaces}, 
J. Chem. Phys. \textbf{100}, 7796 (1994).

\bibitem{muth95} M. Muthukumar, 
\textit{Pattern Recognition by Polyelectrolytes}, 
J. Chem. Phys. \textbf{103}, 4723 (1995).

\bibitem{barr96} J. L. Barrat and J.-F. Joanny, 
\textit{Theory of Polyelectrolyte Solutions}, (Eds: I. Prigogine and S. A. Rice), 
Adv. Chem. Phys. \textbf{94}, 1 (1996).

\bibitem{adam96} Z. Adamczyk and P. Warszynski,
\textit{Role of Electrostatic Interactions in Particle Adsorption},  
Adv. Colloid \& Interf. Sci. \textbf{63}, 41 (1996).

\bibitem{adam03} Z. Adamczyk, 
\textit{Particle Adsorption and Deposition: Role of Electrostatic Interactions}, Adv. Colloid \& Interf. Sci.  
\textbf{100-102}, 267 (2003).

\bibitem{linse96} P. Linse, 
\textit{Adsorption of Weakly Charged Polyelectrolytes at Oppositely Charged Surfaces}, 
Macromolecules \textbf{29}, 326 (1996).

\bibitem{ande98} I. Borukhov and D. Andelman, 
\textit{Scaling Laws of Polyelectrolyte Adsorption}, 
Macromolecules, \textbf{31}, 1665 (1998).

\bibitem{sens99} E. Gurovitch and P. Sens, 
\textit{Adsorption of Polyelectrolyte onto a Colloid of Opposite Charge},
Phys. Rev. Lett. \textbf{82}, 339 (1999).
 
\bibitem{netz99} R. R. Netz and J.-F. Joanny, 
\textit{Adsorption of Semiflexible Polyelectrolytes on Charged Planar Surfaces: Charge Compensation, Charge Reversal, and Multilayer Formation}, 
Macromolecules \textbf{32}, 9013 (1999).

\bibitem{netz99b} R. R. Netz and J.-F. Joanny, 
\textit{Complexation between a Semiflexible Polyelectrolyte and an Oppositely Charged Sphere}, 
Macromolecules \textbf{32}, 9026 (1999).

\bibitem{netz00} K. K. Kunze and R. R. Netz, 
\textit{Salt-Induced DNA-Histone Complexation}, 
Phys. Rev. Lett. \textbf{85}, 4389 (2000).

\bibitem{shaf03} A. Shafir, D. Andelman, and R. R. Netz, 
\textit{Adsorption and Depletion of Polyelectrolytes from Charged Surfaces}, J. Chem. Phys. \textbf{119}, 2355 (2003).

\bibitem{netz14} H. Boroudjerdi, A. Naji, and R. R. Netz, 
\textit{Global Analysis of the Ground-State Wrapping Conformation of a Charged Polymer on an Oppositely Charged Nano-Spheres}, 
Europ. Phys. J. E \textbf{37}, 21 (2014).

\bibitem{kier13} T. A. Kampmann, H. H. Boltz, and J. Kierfeld, 
\textit{Controlling Adsorption of Semiflexible Polymers on Planar and Curved Substrates}, 
J. Chem. Phys. \textbf{139}, 034903 (2013).

\bibitem{fors12} J. Forsman, 
\textit{Polyelectrolyte Adsorption: Electrostatic Mechanisms and Nonmonotonic Responses to Salt Addition}, 
Langmuir \textbf{28}, 5138 (2012).

\bibitem{cher06a} A. G. Cherstvy and R. G. Winkler, 
\textit{Strong and Weak Adsorptions of Polyelectrolyte Chains onto Oppositely Charged Spheres}, 
J. Chem. Phys. \textbf{125}, 064904 (2006).

\bibitem{cher06b} R. G. Winkler and A. G. Cherstvy, 
\textit{Critical Adsorption of Polyelectrolytes onto Charged Spherical Colloids}, Phys. Rev. Lett. \textbf{96}, 066103 (2006).

\bibitem{cher07} R. G. Winkler and A. G. Cherstvy, 
\textit{Adsorption of Weakly Charged Polyelectrolytes onto Oppositely Charged Spherical Colloids}, 
J. Phys. Chem. B \textbf{111}, 8486 (2007).

\bibitem{cher11} A. G. Cherstvy and R. G. Winkler, 
\textit{Polyelectrolyte Adsorption onto Oppositely Charged Interfaces: Unified Approach for Plane, Cylinder, and Sphere}, 
Phys. Chem. Chem. Phys. \textbf{13}, 11686 (2011).

\bibitem{cher12a} A. G. Cherstvy, 
\textit{Critical Polyelectrolyte Adsorption under Confinement: Planar Slit, Cylindrical Pore, and Spherical Cavity}, 
Biopolymers \textbf{97}, 311 (2012).

\bibitem{cher14c} A. G. Cherstvy, 
\textit{Electrostatics and Charge Regulation in Polyelectrolyte Multilayered Assembly}, 
J. Phys. Chem. B \textbf{118}, 4552 (2014).

\bibitem{cher15} S. J. de Carvalho, R. Metzler, and A. G. Cherstvy, 
\textit{Inverted critical Adsorption of Polyelectrolytes in Confinement},
Soft Matter \textbf{11}, 4430 (2015).

\bibitem{cher14} S. J. de Carvalho, R. Metzler, and A. G. Cherstvy, 
\textit{Critical Adsorption of Polyelectrolytes onto Charged Janus Nanospheres}, Phys. Chem. Chem. Phys. \textbf{16}, 15539 (2014).

\bibitem{cher12b} A. G. Cherstvy and R. G. Winkler, 
\textit{Polyelectrolyte Adsorption onto Oppositely Charged Interfaces: Image-Charge Repulsion and Surface Curvature}, 
J. Phys. Chem. B \textbf{116}, 9838 (2012).

\bibitem{dech97} G. Decher, 
\textit{Fuzzy Nanoassemblies: Toward Layered Polymeric Multicomposites},
Science \textbf{277}, 1232 (1997).

\bibitem{schl99} S. T. Dubas and J. B. Schlenoff, 
\textit{ Factors Controlling the Growth of Polyelectrolyte Multilayers}, Macromolecules \textbf{32}, 8153 (1999).

\bibitem{scho03} M. Sch\"onhoff, 
\textit{Self-Assembled Polyelectrolyte Multilayers}, 
Curr. Opin. Colloid \& Interf. Sci. \textbf{8}, 86 (2003).

\bibitem{scho07} M. Sch\"onhoff et al., 
\textit{Hydration and Internal Properties of Polyelectrolyte Multilayers}, Colloids \& Surf. A\textbf{ 303}, 14 (2007).

\bibitem{volo04} D. V. Volodkin, A. I. Petrov, M. Prevot, and G. B. Sukhorukov, \textit{Matrix Polyelectrolyte Microcapsules: New System for Macromolecule Encapsulation}, 
Langmuir \textbf{20}, 3398 (2004).

\bibitem{dub01a} Y. Wang, P. L. Dubin, and H. Zhang, 
\textit{Interaction of DNA with Cationic Micelles:  Effects of Micelle Surface Charge Density, Micelle Shape, and Ionic Strength on Complexation and DNA Collapse}, 
Langmuir \textbf{17}, 1670 (2001).

\bibitem{dub01b} X. H. Feng et al., 
\textit{Critical Conditions for Binding of Dimethyldodecylamine Oxide Micelles to Polyanions of Variable Charge Density}, 
Macromolecules \textbf{34}, 6373 (2001).

\bibitem{dub05}  C. L. Cooper, P. L. Dubin, A. B. Kayitmazer, and S. Turksen, \textit{Polyelectrolyte-Protein Complexes}, 
Curr. Opin. Colloid \& Interf. Sci. \textbf{10}, 52 (2005).

\bibitem{dub06} C. L. Cooper et al., 
\textit{Effects of Polyelectrolyte Chain Stiffness, Charge Mobility, and Charge Sequences on Binding to Proteins and Micelles}, 
Biomacromolecules \textbf{7}, 1025 (2006).

\bibitem{dub07} Y. G. Mishael, P. L. Dubin, R. de Vries, and A. B. Kayitmazer, \textit{Effect of Pore Size on Adsorption of a Polyelectrolyte to Porous Glass},
Langmuir \textbf{23}, 2510 (2007).

\bibitem{dub10} M. Antonov, M. Mazzawi, and P. L. Dubin, 
\textit{Entering and Exiting the Protein-Polyelectrolyte Coacervate Phase via Nonmonotonic Salt Dependence of Critical Conditions}, 
Biomacromolecules \textbf{11}, 51 (2010).

\bibitem{dub11} E. Kizilay, A. B. Kayitmazer, and  P. L. Dubin, 
\textit{Complexation and Coacervation of Polyelectrolytes with Oppositely Charged Colloids}, 
Adv. Colloid \& Interf. Sci. \textbf{167}, 24 (2011).

\bibitem{dub13} A. B. Kayitmazer et al., 
\textit{Protein-Polyelectrolyte Interactions},
Soft Matter \textbf{9}, 2553 (2013).

\bibitem{bork14} I. Szilagyi, G. Trefalt, A. Tiraferri, P. Maroni, and M. Borkovec, 
\textit{Polyelectrolyte Adsorption, Interparticle Forces, and Colloidal Aggregation}, Soft Matter \textbf{10}, 2479 (2014).

\bibitem{bord12} S. Sennato, D. Truzzolillo, and F. Bordi, 
\textit{Aggregation and Stability of Polyelectrolyte-Decorated Liposome Complexes in Water-Salt Media}, 
Soft Matter \textbf{8}, 9384 (2012).

\bibitem{diet10} J.-H. Jeon, J. Adamczik, G. Dietler, and R. Metzler, 
\textit{Denaturation Bubbles in Supercoiled Circular DNA}, 
Phys. Rev. Lett. \textbf{105}, 208101 (2010).

\bibitem{diet12} J. Adamcik, J.-H. Jeon, K. J. Karczewski, R. Metzler, and G. Dietler, 
\textit{Quantifying Supercoiling-induced Denaturation Bubbles in DNA}, 
Soft Matter \textbf{8}, 8651 (2012).

\bibitem{bord15} S. Sennato, L. Carlini, D. Truzzolillo, and F. Bordi, 
\textit{Salt-Induced Reentrant Stability of Polyion-Decorated Particles with tunable Surface Charge Density}, 
Biointerfaces \textbf{137}, 109 (2016).

\bibitem{yu15} S. Yu et al., 
\textit{Interaction of Human Serum Albumin with Short Polyelectrolytes: a Study by Calorimetry and Computer Simulations}, 
Soft Matter \textbf{11}, 4630 (2015).

\bibitem{belt91} S. Beltr\'an, H. H. Hooper, H. W. Blanch, and J. M. Prausnitz, \textit{Monte Carlo Study of Polyelectrolyte Adsorption. Isolated Chains on a Planar Charged Surface},
Macromolecules \textbf{24}, 3178 (1991).

\bibitem{carv10} S. J. de Carvalho, 
\textit{First-order-like Transition in Salt-induced Macroion-Polyelectrolyte Desorption}, 
EPL \textbf{92}, 18001 (2010).

\bibitem{stol02} S. Stoll and P. Chodanowski, 
\textit{Polyelectrolyte Adsorption on an Oppositely Charged Spherical Particle. Chain Rigidity Effects}, 
Macromolecules \textbf{35}, 9556 (2002).

\bibitem{muth02} J. McNamara, C. Y. Kong, and M. Muthukumar, 
\textit{Monte Carlo Studies of Adsorption of a Sequenced Polyelectrolyte to Patterned Surfaces}, J. Chem. Phys. \textbf{117}, 5354 (2002).

\bibitem{stol03} A. Laguecir et al., 
\textit{Interactions of a Polyanion with a Cationic Micelle: Comparison of Monte Carlo Simulations with Experiment}, 
J. Phys. Chem. B \textbf{107}, 8056 (2003).

\bibitem{stol05} S. Ulrich, A. Laguecir, and S. Stoll, 
\textit{Complexation of a Weak Polyelectrolyte with a Charged Nanoparticle. Solution Properties and Polyelectrolyte Stiffness Influences}, 
Macromolecules \textbf{38}, 8939 (2005).

\bibitem{stol06} S. Ulrich, M. Seijo, A. Laguecir, and S. Stoll, 
\textit{Nanoparticle Adsorption on a Weak Polyelectrolyte. Stiffness, pH, Charge Mobility, and Ionic Concentration Effects Investigated by Monte Carlo Simulations}, J. Phys. Chem. B \textbf{110}, 20954 (2006).

\bibitem{dubi10} A. B. Kayitmazer, B. Quinn, K. Kimura, G. L. Ryan, A. J. Tate, D. A. Pink and P. L. Dubin, 
\textit{Protein Specificity of Charged Sequences in Polyanions and Heparins},
Biomacromolecules \textbf{11}, 3325 (2010).

\bibitem{stol11} S. Ulrich, M. Seijo, F. Carnal, and S. Stoll, 
\textit{Formation of Complexes between Nanoparticles and Weak Polyampholyte Chains. Monte Carlo Simulations},
Macromolecules \textbf{44}, 1661 (2011).

\bibitem{carv14} V. M. de Oliveira and S. J. de Carvalho, 
\textit{Adsorption of pH-responsive Polyelectrolyte Chains onto Spherical Macroions}, 
Eur. Phys. J. E  \textbf{37}, 75 (2014). 

\bibitem{stol09} M. Seijo, M. Pohl, S. Ulrich, and S. Stoll, 
\textit{Dielectric Discontinuity Effects on the Adsorption of a Linear Polyelectrolyte at the Surface of a Neutral Nanoparticle}, 
J. Chem. Phys. \textbf{131}, 174704 (2009).

\bibitem{mess09} R. Messina, 
\textit{Electrostatics in Soft Matter}, 
J. Physics: Condens. Matt. \textbf{21}, 113102 (2009).

\bibitem{dobr07} J.-M. Y. Carrillo and A. V. Dobrynin, 
\textit{Molecular Dynamics Simulations of Polyelectrolyte Adsorption}, 
Langmuir \textbf{23}, 2472 (2007).

\bibitem{hoda08} N. Hoda and S. Kumar, 
\textit{Theory of Polyelectrolyte Adsorption onto Surfaces Patterned with Charge and Topography}, 
J. Chem. Phys. \textbf{128}, 124907 (2008).

\bibitem{hoda07} N. Hoda and S. Kumar, 
\textit{Brownian Dynamics Simulations of Polyelectrolyte Adsorption in Shear Flow with Hydrodynamic Interaction}, 
J. Chem. Phys. \textbf{127}, 234902 (2007).

\bibitem{vatt08} S. V. Lyulin, I. Vattulainen, and A. A. Gurtovenko, 
\textit{Complexes Comprised of Charged Dendrimers, Linear Polyelectrolytes, and Counterions: Insight through Coarse-Grained Molecular Dynamics Simulations}, Macromolecules \textbf{41}, 4961 (2008).

\bibitem{muth11} J. Wang and M. Muthukumar, 
\textit{Encapsulation of a Polyelectrolyte Chain by an Oppositely Charged Spherical Surface}, 
J. Chem. Phys. \textbf{135}, 194901 (2011).

\bibitem{chin11} Z. Wang et al., 
\textit{Charge Inversion by Flexible Polyelectrolytes on Spherical Surfaces: Numerical Self-Consistent Field Calculations under the Ground-State Dominance Approximation}, 
Macromolecules \textbf{44}, 8607 (2011).

\bibitem{vrie04} R. de Vries, 
\textit{Monte Carlo Simulations of flexible Polyanions complexing with Whey Proteins at their Isoelectric Point}, 
J. Chem. Phys. \textbf{120}, 3475 (2004). 

\bibitem{kong98} C. Y. Kong and M. Muthukumar, 
\textit{Monte-Carlo Study of Adsorption of a Polyelectrolyte onto Charged Surfaces}, 
J. Chem. Phys. \textbf{109}, 1522 (1998).

\bibitem{krama96} E. Y. Kramarenko et al., 
\textit{Molecular Dynamics Simulation Study of Adsorption of Polymer Chains with variable Degree of Rigidity. I. Static Properties}, 
J. Chem. Phys. \textbf{104}, 4806 (1996).

\bibitem{mess06} R. Messina, 
\textit{Effect of Image Forces on Polyelectrolyte Adsorption at a Charged Surface}, 
Phys. Rev. E \textbf{70}, 051802 (2004).

\bibitem{reddy06} G. Reddy, R. Chang, and A. Yethiraj, 
\textit{Adsorption and Dynamics of a Single Polyelectrolyte Chain near a Planar Charged Surface:? Molecular Dynamics Simulations with Explicit Solvent},
J. Chem. Theor. Comput. \textbf{2}, 630 (2006).

\bibitem{caet13} S. J. de Carvalho and D. L. Z. Caetando, 
\textit{Adsorption of Polyelectrolytes onto Oppositely Charged Cylindrical Macroions}, 
J. Chem. Phys. \textbf{138}, 244909 (2013).

\bibitem{moli14} G. Luque-Caballero, A. Martin-Molina, and M. Quesada-Perez, \textit{Polyelectrolyte Adsorption onto Like-Charged Surfaces mediated by Trivalent Counterions: A Monte Carlo Simulation Study}, 
J. Chem. Phys. \textbf{140}, 174701 (2014).

\bibitem{bach15} J. Gross, T. Vogel, and M. Bachmann, 
\textit{Structural Phases of Adsorption for Flexible Polymers on Nanocylinder Surfaces}, 
Phys. Chem. Chem. Phys. \textbf{17}, 3070 (2015).

\bibitem{lind91} T. Lindstrom, J. C. Roberts (Ed.), p. 25 in "Paper Chemistry", (Blackie, London, 1991).

\bibitem{moeh06} D. G. Shchukin, M. Zheludkevich, K. Yasakau, S. Lamaka, M. G. S. Ferreira, and H. M\"ohwald, 
\textit{Layer-by-Layer Assembled Nanocontainers for Self-Healing Corrosion Protection},
Adv. Mater. \textbf{18}, 1672 (2006).

\bibitem{andr10} D. V. Andreeva, E. V. Skorb, and D. G. Shchukin, 
\textit{Layer-by-Layer Polyelectrolyte/Inhibitor Nanostructures for Metal Corrosion Protection},
ACS Appl. Mater. Interfaces \textbf{2}, 1954 (2010).

\bibitem{napp83} D. H. Napper, "Polymeric Stabilization of Colloidal Dispersions", (Academic Press, New York, 1983).

\bibitem{pinc91} P. Pincus, 
\textit{Colloid Stabilization with Grafted Polyelectrolytes},
 Macromolecules \textbf{24}, 2912 (1991).

\bibitem{pavi08} S. Pavlidou and C. D. Papaspyrides, 
\textit{A Review on Polymer-layered Silicate Nanocomposites}, 
Prog. Polym. Sci. \textbf{33}, 1119 (2008).

\bibitem{golu01} A. K. Chakraborty and A. J. Golumbfskie, 
\textit{Polymer Adsorption-Driven Self-Assembly of Nanostructures}, 
Annu. Rev. Phys. Chem. \textbf{52}, 537 (2001).

\bibitem{malw03} G. Schmidt and M. M. Malwitz, 
\textit{Properties of Polymer-Nanoparticle Composites}, 
Curr. Opin. Colloid \& Interf. Sci. \textbf{8}, 103 (2003).

\bibitem{ball13} N. Welsch, Y. Lu, J. Dzubiella, and M. Ballauff, 
\textit{Adsorption of Proteins to Functional Polymeric Nanoparticles},
Polymer \textbf{54}, 2835 (2013).

\bibitem{dubi94} "Macromolecular Complexes in Chemistry and Biology", Eds.: P. L. Dubin, J. Bock, R. M. Davis, D. Schulz, and C. Thies (Springer Verlag, Berlin, 1994).

\bibitem{greg01} S.-K. Kam and J. Gregory, 
\textit{The Interaction of Humic Substances with Cationic Polyelectrolytes},
Water Res. \textbf{35}, 3557 (2001).

\bibitem{greg11} J. Gregory and S. Barany, 
\textit{Adsorption and Flocculation by Polymers and Polymer Mixtures},
 Adv. Colloid \& Interf. Sci. \textbf{169}, 1 (2011).

\bibitem{borg14} J. Borges and J. F. Mano, 
\textit{Molecular Interactions Driving the Layer-by-Layer Assembly of Multilayers},
Chem. Rev. \textbf{114}, 8883 (2014).

\bibitem{gees09} B. G. De Geest et al., 
\textit{Polyelectrolyte Microcapsules for Biomedical Applications}, 
Soft Matter \textbf{5}, 282 (2009).

\bibitem{skir11} M. Delcea, H. M\"ohwald, and A. G. Skirtach, 
\textit{Stimuli-Responsive LbL Capsules and Nanoshells for Drug Delivery},
Adv. Drug Deliv. Rev. \textbf{63}, 730 (2011).


\bibitem{cohe10} M. A. Cohen Stuart et al., 
\textit{Emerging Applications of Stimuli-Responsive Polymer Materials}, 
Nature Mat. \textbf{9}, 101 (2010).

\bibitem{lasc00} P. Bertrand, A. Jonas, A. Laschewsky, R. Legras, 
\textit{Ultrathin Polymer Coatings by Complexation of Polyelectrolytes at Interfaces: Suitable Materials, Structure and Properties}, 
Macromol. Rapid Commun. \textbf{21}, 319 (2000).

\bibitem{boud09} T. Boudou, T. Crouzier, K. Ren, G. Blin, and C. Picart, \textit{Multiple Functionalities of Polyelectrolyte Multilayer Films: New Biomedical Applications}, 
Adv. Mater. \textbf{22}, 441 (2010).

\bibitem{cull04} T. M. Allen and P. R. Cullis,
\textit{Drug Delivery Systems: Entering the Mainstream},
Science \textbf{303}, 1818 (2004).

\bibitem{gomp15} A. Wysocki, J. Elgeti, and G. Gompper, 
\textit{Giant Adsorption of Microswimmers: Duality of Shape Asymmetry and Wall Curvature}, 
Phys. Rev. E \textbf{91}, 050302 (2015).

\bibitem{eise93} E. Eisenriegler, "Polymers near Interfaces", (World Scientific, Singapore, 1993).

\bibitem{eise82} E. Eisenriegler, K. Kremer, and K. Binder, 
\textit{Adsorption of Polymer Chains at Surfaces: Scaling and Monte Carlo Analyses}, 
J. Chem. Phys. \textbf{77}, 6296 (1982).

\bibitem{eise96} E. Eisenriegler, A. Hanke, and S. Dietrich, 
\textit{Polymers Interacting with Spherical and Rodlike Particles}, 
Phys. Rev. E \textbf{54}, 1134 (1996).

\bibitem{ohsh82} H. Ohshima, T. W. Healy, and L. R. White, 
\textit{Accurate Analytic Expressions for the Surface Charge Density/Surface Potential Relationship and Double-Layer Potential Distribution for a Spherical Colloidal Particle}, 
J. Colloid \& Interf. Sci. \textbf{90}, 17 (1982).

\bibitem{ohsh98} H. Ohshima, 
\textit{Surface Charge Density/Surface Potential Relationship for a Cylindrical Particle in an Electrolyte Solution}, 
J. Colloid \& Interf. Sci. \textbf{200}, 291 (1998).

\bibitem{ande95} D. Andelman, \textit{"Electrostatic Properties of Membranes: the Poisson-Boltzmann Theory"}, Ch. 12, p. 603, in "Handbook of Biological Physics", Eds.: R. Lipowsky and E. Sackmann, (Elsevier, 1995).

\bibitem{kapp06} H.-J. Butt, K. Graf, and M. Kappl, " Physics and Chemistry of Interfaces", (Wiley, 2006).

\bibitem{anti98} A. M. O. Mohamed and H. E. Antia, 
\textit{Geoenvironmental Engineering}, Ch. 5.6.2, (Elsevier, 1998).



\bibitem{fren10} R. H. French et al., 
\textit{ Long Range Interactions in Nanoscale Science}, 
Rev. Mod. Phys. \textbf{82}, 1887 (2010).

\bibitem{korn07} A. A. Kornyshev, D. J. Lee, S. Leikin, and A. Wynveen, 
\textit{Structure and Interactions of Biological Helices}, 
Rev. Mod. Phys. \textbf{79}, 943 (2007).

\bibitem{podg09} D. Ben-Yaakov, D. Andelman, D. Harries, and R. Podgornik,
\textit{Beyond Standard Poisson-Boltzmann Theory: Ion-specific Interactions in Aqueous Solutions}, 
J. Phys.: Condens. Matter \textbf{21}, 424106 (2009).

\bibitem{podg13} A. Naji, M. Kanduc, J. Forsman, and R. Podgornik, 
\textit{Perspective: Coulomb Fluids--Weak Coupling, Strong Coupling, in between and beyond}, 
J. Chem. Phys. \textbf{139}, 150901 (2013).

\bibitem{levy12} A. Levy, D. Andelman, and H. Orland,
\textit{Dielectric Constant of Ionic Solutions: A Field-Theory Approach},
Phys. Rev. Lett. \textbf{108}, 227801 (2012).

\bibitem{roij12} M. M. Hatlo, R. van Roij, and L. Lue, 
\textit{ The Electric Double Layer at High Surface Potentials: The Influence of Excess Ion Polarizability},
Europhys. Lett. \textbf{97}, 28010 (2012).


\bibitem{oosa70} F. Oosawa, \textit{"Polyelectrolytes"}, (Marcel Dekker, New York, 1971).

\bibitem{mann78} G. S. Manning, 
\textit{The Molecular Theory of Polyelectrolyte Solutions with Applications to the Electrostatic Properties of Polynucleotides}, 
Quart. Rev. Biophys. \textbf{11}, 179 (1978).

\bibitem{netz02} A. G. Moreira and R. R. Netz, 
\textit{Simulations of Counterions at Charged Plates}, 
Eur. Phys. J. E \textbf{8}, 33 (2002).

\bibitem{gros00} A. Yu. Grosberg, T. T. Nguyen, and B. I. Shklovskii, 
\textit{Colloquium: The Physics of Charge Inversion in Chemical and Biological Systems}, 
Rev. Mod. Phys. \textbf{74}, 329 (2002).

\bibitem{netz05} H. Boroudjerdi, Y.-W. Kim, A. Naji, R. R. Netz, X. Schlagberger, and A. Serr,
\textit{Statics and Dynamics of Strongly Charged Soft Matter},
Phys. Rep. \textbf{416}, 129 (2005).

\bibitem{pars00} W. M. Gelbart, R. F. Bruinsma, P. A. Pincus, and V. A. Parsegian, \textit{DNA-Inspired Electrostatics}, 
Phys. Today \textbf{53}, 38 (2000).

\bibitem{levi02} Y. Levin, 
\textit{Electrostatic Correlations: from Plasma to Biology}, 
Rep. Prog. Phys. \textbf{65}, 1577 (2002).

\bibitem{wong03} T. E. Angelini, H. Liang, W. Wriggers‡, and G. C. L. Wong, \textit{Likecharge Attraction between Polyelectrolytes induced by Counterion Charge Density Waves}, 
Proc. Natl. Acad. Sci. U. S. A. \textbf{100}, 8634 (2003).

\bibitem{eige59} M. Eigen and E. Wicke, 
\textit{The Thermodynamics of Electrolytes at Higher Concentration}, 
J. Phys. Chem. \textbf{58}, 702 (1954).

\bibitem{ande97} I. Borukhov, D. Andelman, and H. Orland, 
\textit{Steric Effects in Electrolytes: A Modified Poisson-Boltzmann Equation}, Phys. Rev. Lett. \textbf{79}, 435 (1997).

\bibitem{ande06} D. Andelman, in \textit{"Soft Condensed Matter Physics in Molecular and Cell Biology" (Scottish Graduate Series)}, 1st edition, Eds: W. C. K. Poon and D. Andelman, (Taylor \& Francis, CRC Press, 2006).
 
\bibitem{korn07b} A. A. Kornyshev, \textit{Double-Layer in Ionic Liquids: Paradigm Change?}, J. Phys. Chem. B \textbf{111}, 5545 (2007).

\bibitem{onsa33} L. Onsager, 
\textit{Theories of Concentrated Electrolytes},
Chem. Rev. \textbf{13}, 73 (1933).

\bibitem{dese00} M. Deserno, PhD Thesis, 
\textit{"Counterion Condensation for Rigid Linear Polyelectrolytes"}, Mainz, Germany, (2000).

\bibitem{soka88} N. Madras and  A. D. Sokal, 
\textit{The Pivot Algorithm:  A Highly  Efficient Monte  Carlo Method  for  the  Self-Avoiding  Walk}, 
J. Stat. Phys. \textbf{50}, 109 (1988).

\bibitem{labb05} B. J\"onsson, A. Nonat, C. Labbez, B. Cabane, and H. Wennerstr\"om, \textit{Controlling the Cohesion of Cement Paste}, 
Langmuir \textbf{21}, 9211 (2005).

\bibitem{labb14} M. Turesson, A. Nonat, and C. Labbez, 
\textit{Stability of Negatively Charged Platelets in Calcium-Rich Anionic Copolymer Solutions}, 
Langmuir \textbf{30}, 6713 (2014).

\bibitem{cher11b} A. G. Cherstvy, 
\textit{Electrostatic Interactions in Biological DNA-related Systems}, 
Phys. Chem. Chem. Phys. \textbf{13}, 9942 (2011).


\bibitem{blaa90} J. Blaakmeer, M. R. Bohmer, M. A. Cohen Stuart, and G. J. Fleer, 
\textit{Adsorption of Weak Polyelectrolytes on Highly Charged Surfaces. Poly(acrylic acid) on Polystyrene Latex with Strong Cationic Groups}, 
Macromolecules \textbf{23}, 2301 (1990).

\bibitem{atal14} S. Atalay, Y. Ma, and S. Qian, 
\textit{Analytical Model for Charge Properties of Silica Particles},
J. Colloid \& Interf. Sci. \textbf{425}, 128 (2014).

\bibitem{zsom87} W. M. Brouwer and R. L. J. Zsom, 
\textit{Polystyrene Latex Particle Surface Characteristics},
Colloids \& Surf. \textbf{24}, 195 (1987). 

\bibitem{roja02} O. J. Rojas, 
\textit{Adsorption of Polyelectrolytes on Mica}, in "Encyclopedia of Surface and Colloid Science", p. 517,  (Marcel Dekker, 2002).

\bibitem{barl86} D. J. Barlow and J. M. Thornton, \textit{The Distribution of Charged Groups in Proteins}, 
Biopolymers \textbf{25}, 1717 (1986).

\bibitem{volo14} D. Volodkin, R. von Klitzing, and H. M\"ohwald, 
\textit{Polyelectrolyte Multilayers: Towards Single Cell Studies}, 
Polymers \textbf{6}, 1502 (2014).

\bibitem{pogh13} A. Poghossian, M. Weil, A. G. Cherstvy, and M. J. Sch\"oning, \textit{Electrical Monitoring of Polyelectrolyte Multilayer Formation by Means of Capacitive Field-Effect Devices}, 
Analyt. Bioanalyt. Chem. \textbf{405}, 6425 (2013).

\bibitem{peyr04} C. S. Peyratout and L. D\"ahne, 
\textit{Tailor-Made Polyelectrolyte Microcapsules: From Multilayers to Smart Containers}, 
Angew. Chem. Intl. Ed. Engl. \textbf{43}, 3762 (2004).

\bibitem{hamm99} P. T. Hammond, 
\textit{Recent Explorations in Electrostatic Multilayer Thin Film Assembly}, Curr. Opin. Colloid \& Interf. Sci. \textbf{4}, 430 (1999).

\bibitem{dech12} G. Decher and J. B. Schlenoff (Eds.), \textit{Multilayer Thin Films: Sequential Assembly of Nanocomposite Materials}, 2nd ed., (Wiley, 2012).

\bibitem{imot10} T. Imoto, T. Kida, M. Matsusaki, and M. Akashi, 
\textit{Preparation and Unique pH-Responsive Properties of Novel Biodegradable Nanocapsules Composed of Poly($\gamma$-glutamic acid) and Chitosan as Weak Polyelectrolytes}, 
Macromol. Biosci. \textbf{10}, 271 (2010).

\bibitem{abba16} M. A. Brown, A. Goel, and Z. Abbas, 
\textit{Effect of Electrolyte Concentration on the Stern Layer Thickness at a Charged Interface}, 
Angew. Chem. Int. Ed. Engl. \textbf{55}, 3790 (2016).

\bibitem{bari14} M. Barisik, S. Atalay, A. Beskok, and S. Qian, 
\textit{Size Dependent Surface Charge Properties of Silica Nanoparticles}, J. Phys. Chem. C \textbf{118}, 1836 (2014).


\bibitem{schu97} V. Shubin, 
\textit{Adsorption of Cationic Polyacrylamide onto Monodisperse Colloidal Silica from Aqueous Electrolyte Solutions}, 
J. Colloid \& Interf. Sci. \textbf{191}, 372 (1997).
 
\bibitem{lind12} F. Cuomo et al., 
\textit{pH-responsive Liposome-Templated Polyelectrolyte Nanocapsules}, 
Soft Matter \textbf{8}, 4415 (2012).

\bibitem{hube14} S. T. Moerz and P. Huber, 
\textit{Protein Adsorption into Mesopores: A Combination of Electrostatic Interaction, Counterion Release, and van der Waals Forces}, 
Langmuir \textbf{30}, 2729 (2014).

\bibitem{hube15a} S. T. Moerz and P. Huber, 
\textit{pH-Dependent Selective Protein Adsorption into Mesoporous Silica}, J. Phys. Chem. C \textbf{119}, 27072 (2015).

\bibitem{hube15b} P. Huber, 
\textit{Soft Matter in Hard Confinement: Phase Transition Thermodynamics, Structure, Texture, Diffusion and flow in nanoporous media}, 
J. Phys.: Condens. Matt. \textbf{27}, 103102 (2015).

\bibitem{lifs51} R. M. Fuoss, A. Katchalsky and S. Lifson, 
\textit{The Potential of an Infinite Rod-Like Molecule and the Distribution of the Counter Ions}, 
Proc. Natl. Acad. Sci. U. S. A. \textbf{37}, 579 (1951).

\end{thebibliography}
\end{document}